\documentclass{article}
\usepackage[a4paper,left=2cm,right=3cm,top=3cm,bottom=2cm]{geometry}
\usepackage{url} 
\usepackage{lineno}
\usepackage{soul}
\usepackage[utf8]{inputenc}
\usepackage[english]{babel}
\usepackage{amsmath}
\usepackage{amsfonts}
\usepackage{amssymb}
\usepackage{graphicx}
\usepackage[round]{natbib}
\usepackage{hyperref}
\usepackage{enumitem}
\usepackage{hyperref, newfloat, caption, subcaption}
\hypersetup{
	colorlinks = true,
	urlcolor=blue,
	linkcolor=blue,
	citecolor=blue
}

\graphicspath{{Figures/DEF/}}

\newcommand{\para}{\boldsymbol{\theta} }

\newcommand{\data}{\boldsymbol{y} }

\newcommand{\dataRV}{\boldsymbol{Y} }

\newcommand{\datanoise}{\boldsymbol{\varepsilon_{\mathcal{O}}}}

\newcommand{\lat}{\boldsymbol{X} }

\newcommand{\ssmc}{PostRisk-SMC }
\newcommand{\ssmcbracket}{(PostRisk-SMC) }

\usepackage{algpseudocode}
\usepackage{algorithm}
\usepackage{multicol}
\usepackage{bbm}

\usepackage{tikz}
\usetikzlibrary{shapes,arrows}
\usetikzlibrary{positioning}

\usepackage{graphicx}
\usepackage{amsmath}
\usepackage{mathptmx} 
\usepackage{setspace}

\usepackage{color, colortbl}
\definecolor{LightCyan}{rgb}{0.88,1,1}

\begin{document}

\LARGE{\textbf{Rare event probability estimation for groundwater inverse \\ problems with a two-stage Sequential Monte Carlo approach}} \\

\large{Lea Friedli$^{1,2}$ and Niklas Linde$^1$}\\

$^1$ Institute of Earth Sciences, University of Lausanne, Switzerland\\
$^2$ Institute of Mathematical Statistics and Actuarial Science, University of Bern, Switzerland \\

\begin{abstract}
Bayesian inversions followed by estimations of rare event probabilities are often needed to analyse groundwater hazards. Instead of focusing on the posterior distribution of model parameters, the main interest lies then in the distribution of a specific quantity of interest contingent upon these parameters. To address the associated methodological challenges, we introduce a two-stage Sequential Monte Carlo approach. In the first stage, it generates particles that approximate the posterior distribution; in the second stage, it employs subset sampling techniques to assess the probability of the rare event of interest. By considering two hydrogeological problems of increasing complexity, we showcase the efficiency and accuracy of the resulting \ssmc method for rare event probability estimation related to groundwater hazards. We compare the performance of the \ssmc method with a traditional Monte Carlo approach that relies on Markov chain Monte Carlo samples. We showcase that our estimates align with those of the traditional method, but the coefficients of variation are notably lower for the same computational budget when targeting more rare events. Furthermore, we highlight that the \ssmc method allows estimating rare event probabilities approaching one in a billion using less than one hundred thousand forward simulations. Even if the presented examples are related to groundwater hazards, the methodology is well-suited for addressing a wide range of topics in the geosciences and beyond.
\end{abstract}

\section{Introduction}
Decision-making processes concerning groundwater and other environmental systems are subject to uncertainty. Consequently, decision-making often involves the identification and avoidance of hazards while assessing associated risks. While a hazard represents a dangerous phenomenon itself, risk considers the resulting potential of harm for human individuals or economic assets \citep{ward2020natural}. Risk assessment plays a crucial role in the context of groundwater management, as fresh and uncontaminated groundwater is a prerequisite for global water security \citep{famiglietti2014global} and as remediation of contaminated aquifers is extremely costly and time-consuming. Groundwater contamination and over-exploitation have not only direct adverse consequences for humans, but also for ecosystems and ecosystem services.  \par

Our focus is on a particular aspect of risk assessment, namely, the estimation of the probability of a hazard occurring. This hazard is defined by a quantity of interest that takes on critical values. For a precise analysis of hazard occurrence, it is essential to consider the uncertainty associated with the parameters of a conceptual model. Hence, in the field of hydrogeology, Monte Carlo approaches for sampling uncertain hydrological model parameters have been widely employed (e.g., \citeauthor{lahkim1999stochastic} \citeyear{lahkim1999stochastic}, \citeauthor{khadam2003multi} \citeyear{khadam2003multi}, \citeauthor{benekos2007probabilistic} \citeyear{benekos2007probabilistic}, \citeauthor{siirila2012quantitative} \citeyear{siirila2012quantitative},  \citeauthor{enzenhoefer2012probabilistic} \citeyear{enzenhoefer2012probabilistic}). Such approaches can be challenging to apply in practice since hazards often fall under the category of rare events, requiring specialized modeling techniques to accurately represent the tail behavior of the quantity of interest. In this context, classical Monte Carlo estimation is impractical as it requires an excessively large sample (\citeauthor{cerou2012sequential} \citeyear{cerou2012sequential}). One approach to mitigate the computational burden is to combine Monte Carlo methods with surrogate modeling (e.g., \citeauthor{li2010evaluation} \citeyear{li2010evaluation}), thereby speeding up the computation time of forward evaluations. Another option is to employ importance sampling in order to focus the sampling on critical regions of the quantity of interest. However, selecting a well-working importance density for high-dimensional problems is often difficult \citep{au2003important}. Extreme value theory (e.g., \citeauthor{brodin2008extreme} \citeyear{brodin2008extreme}), relying on fitting an extreme value distribution to represent the distribution of the quantity of interest, offers yet another alternative and is widely used to predict probabilities of environmental hazards such as extreme floods \citep{morrison2002stochastic}. Extreme value theory necessitates sizable sample sizes for distribution fitting, is contingent on the chosen distribution's shape, and does not offer simulations of the rare event (e.g., \citeauthor{diebold2000pitfalls} \citeyear{diebold2000pitfalls}). An alternative data-intensive method for estimating extremes is based on the 'Peaks over Threshold' technique (POT; \citeauthor{leadbetter1991basis} \citeyear{leadbetter1991basis}). In this approach, extreme events are analysed by focusing on values that exceed a certain threshold. \par

% risk and inversion
We perform rare event probability estimation for the case when indirect site-specific data~$\data$ are available (e.g., from tracer or pumping tests). We employ a Bayesian framework in which the hydrogeological parameters $\para$ are characterized by a  posterior probability density function (PDF) $p(\para | \data)$, given by the distribution of $\para$ conditioned on measurements $\data$. Compared to a standard Bayesian inversion problem in which the end-product is an approximation of the posterior PDF, we interrogate the distribution of a quantity of interest depending on the parameters through a non-linear relationship $\para \mapsto \mathcal{R}(\para)$, for instance, in the probability of this quantity exceeding a critical threshold (considered as the hazard). In practical scenarios, the presence of non-linearity frequently precludes the availability of an analytical formula for the distribution of the quantity of interest when conditioned on the data $\data$. \par

In structural engineering, similar problems have been addressed by performing probabilistic updating of system parameters using dynamic data and subsequently updating the estimation of the system's reliability (e.g., \citeauthor{papadimitriou2001updating}, \citeyear{papadimitriou2001updating}). In this context, \citet{straub2011reliability} introduced
the so-called Bayesian Updating with Structural reliability methods (BUS; e.g. \citeauthor{straub2015bayesian} \citeyear{straub2015bayesian}). For the Bayesian analysis, BUS can be interpreted as an extension of rejection sampling (\citeauthor{ripley2009stochastic} \citeyear{ripley2009stochastic}). To extend BUS for posterior rare event probability estimation, \citet{straub2016bayesian} present an approach targeting both the posterior and the rare event by using reliability methods. A challenge of this method is the selection of the constant employed in the extended rejection sampling, as its choice can impact overall performance. In a similar approach targeting `updated robust reliability measures', \citet{jensen2013use} rely on transitional MCMC \citep{ching2007transitional} to derive a set of posterior samples followed by subset sampling for the reliability analysis. A very different approach enabling the combination of inference and rare event estimation that has been explored in the geosciences is Bayesian Evidential Learning (BEL; \citeauthor{hermans2016direct} \citeyear{hermans2016direct}), which aims to learn a direct relationship between measurements and quantity of interest by sampling from the prior distribution (e.g., \citeauthor{thibaut2021new} \citeyear{thibaut2021new}). For higher-dimensional parameter spaces and non-linear relationships, it can be difficult for BEL to capture the full joint distribution with a reasonable number of samples. \par

We propose a two-stage application of Sequential Monte Carlo (SMC; \citeauthor{doucet2001introduction} \citeyear{doucet2001introduction}), which we refer to as the Posterior Risk Sequential Monte Carlo \ssmcbracket method. Bayesian inversion in hydrogeology and other environmental fields is often addressed using Markov chain Monte Carlo (MCMC) methods. For high-dimensional problems with non-linear forward solvers, standard MCMC methods often have difficulties in approximating the posterior PDF within realistic computational constraints. This happens as the Markov chains may be trapped in local minima for long times or have insignificant probabilities of switching between posterior modes (e.g., \citeauthor{neal2001annealed} \citeyear{neal2001annealed}, \citeauthor{amaya2022hydrogeological} \citeyear{amaya2022hydrogeological}). To overcome these challenges, methods based on so-called power posteriors have been introduced. In a power posterior, the likelihood function is downweighted by exponentiating it with the inverse of a temperature greater than one, a process known as tempering. This facilitates more straightforward exploration at higher temperatures. Parallel tempering \citep{earl2005parallel} is an MCMC approach where interacting chains target different power posteriors, allowing states sampled at higher temperatures to propose states in chains targeting the posterior distribution. In geophysical inversion, \citet{sambridge2014parallel} illustrated that parallel tempering significantly enhances sampling efficiency, enabling a more extensive exploration of the parameter space in comparison to conventional MCMC methods. Similar to traditional MCMC techniques, parallel tempering approximates the posterior using states sampled post burn-in. In contrast, annealed importance sampling \citep{neal2001annealed} relies on sequential importance sampling. While in MCMC, the accuracy of posterior estimates relies on the precise identification of the burn-in period for the chains, annealed importance sampling ensures asymptotic correctness through importance steps, even when errors occur in approximating intermediate distributions \citep{neal2001annealed}. The SMC method \citep{doucet2001introduction} is based on annealed importance sampling. As a particle method, it provides a weighted sample of particles for posterior approximation by simulating a sequence of power posteriors transferring the prior PDF to the posterior PDF by successively increasing the weight of the likelihood (\citeauthor{del2007sequential} \citeyear{del2007sequential}). While the SMC method is extensively used in science and engineering, it has only seen limited use in the geosciences (i.e., \citeauthor{vrugt2013hydrologic} \citeyear{vrugt2013hydrologic}, \citeauthor{linde2017uncertainty} \citeyear{linde2017uncertainty}). We build our \ssmc method on an adaptive version of the SMC method by \citet{zhou2016toward}, which automatically tunes the cooling sequence between power posteriors. Recently, adaptive SMC methods have been employed successfully for geophysical \citep{amaya2021adaptive, davies2021bayesian} and hydrogeological \citep{amaya2022hydrogeological} inversion problems, demonstrating superior performance compared with state-of-the-art MCMC methods. \par

Relying only on a particle approximation of the posterior PDF is insufficient when estimating rare event probabilities. As a relatively small number (tens or hundreds, sometimes thousands) of particles is used in practice, this means that no particle is likely to be associated with the rare event that might, for instance, have a probability of one in a million. To address this, a new SMC formulation has emerged that specifically targets rare events by employing a sequence of nested sets pertaining to the hazard scenario. This sequence refers to a hierarchical structure of sets with each set being a subset of the set above it. In a scenario targeting the probability of the quantity of interest exceeding a critical threshold, the nested sets are related to intervals $[T_k, \infty)$ with thresholds $T_k$ increasing from minus infinity to the threshold of interest. This approach relies on the fact that the small probability of the rare event can be expressed as a product of larger conditional probabilities involving the intermediate nested sets. Such a splitting technique was first introduced as `subset sampling' by \citet{au2001estimation} in the context of reliability analysis and has been applied, for instance, in the context of radioactive waste management (e.g., \citeauthor{cadini2012subset} \citeyear{cadini2012subset}) and earthquake engineering (e.g., \citeauthor{au2003subset} \citeyear{au2003subset}). In the SMC literature, subset sampling is presented by \citet{del2006sequential} and \citet{johansen2005sequential}. \citet{cerou2012sequential} and \citet{botev2008efficient} extended the existing methods by using an adaptive method that optimally selects the subsets on the fly. Subset sampling has been further leveraged by employing surrogates \citep{bourinet2011assessing} or by employing a multilevel approach \citep{ullmann2015multilevel}. While all of these applications rely on uncertain parameters $\para$ following a `prior' PDF, we here adapt this approach to rare event estimation with respect to a posterior PDF that is first approximated by adaptive SMC. The resulting \ssmc method relies on the same principles as the approach of \citet{jensen2013use} but within the theoretical formulation of particle methods and SMC. Further, \citet{jensen2013use} consider engineering applications and dynamic data, while we introduce the \ssmc in the context of hydrogeological rare event probability estimation. In addition, we perform resampling of the particles only occasionally (during the posterior phase), while the transitional MCMC approach applied by \citet{jensen2013use} does so in every iteration. Since resampling impacts the variance of estimates \citep{douc2005comparison}, it is usually beneficial to resample only when the variation in the particle weights becomes too high. \par

For comparison purposes, we consider a conventional Monte Carlo approach for the rare event probability estimation, as applied for instance by \citet{dall2023probabilistic} for risk assessment of groundwater inflow in the context of tunnel construction. In our inversion setting, we rely on MCMC samples approximating the posterior PDF for the Monte Carlo estimation. Our first example consists of a simplified one-dimensional flow scenario where we utilize pumping tests to estimate the probability of high flow rates. Subsequently, we consider a more realistic two-dimensional flow and transport problem, focusing on assessing the probability of contamination breakthrough. The remainder of the manuscript is organized as follows: Section \ref{SMC_ssmc_metho} gives a methodological overview of the considered setting and introduces the \ssmc method; Section \ref{SMC_ssmc_1d} presents the one-dimensional flow example and Section \ref{SMC_ssmc_2d} the two-dimensional transport example; finally, the study ends with a discussion and conclusions in Sections \ref{SMC_ssmc_disc} and~\ref{SMC_ssmc_conc}, respectively.

\section{Methodology}
\label{SMC_ssmc_metho}

\subsection{Notation}

We target an unknown property vector $\para \in\mathbb{R}^P$ representing a model domain from which we obtain measurements $\data~\in~\mathbb{R}^M$. We consider a setting where measurements are realizations of the random variable $\dataRV = \mathcal{G}(\para) + \datanoise$, with $\mathcal{G}: \mathbb{R}^P \rightarrow \mathbb{R}^M$ referring to the forward operator and $\datanoise$ to the observational noise. Assuming independent Gaussian observational errors, we express the likelihood as $p(\data | \para)=\varphi_{M}(\data; \mathcal{G}(\para), \boldsymbol{\Sigma_{\dataRV}})$, with $\varphi_{M}(\cdot; \mathcal{G}(\para), \boldsymbol{\Sigma_{\dataRV}})$ denoting the PDF of a $M$-variate normal distribution with the mean $\mathcal{G}(\para)$ and the diagonal covariance matrix $\boldsymbol{\Sigma_{\dataRV}}$ of the observational errors. While we have opted for the simplicity of assuming independent Gaussian observational errors, the methodology remains applicable in a broader context, accommodating alternative error assumptions.   \par

We consider a quantity of interest $R = \mathcal{R}(\para)$ derived from $\para$ via some function $\mathcal{R}: \mathbb{R}^P \rightarrow \mathbb{R}$. More specifically, we target a rare set $A = \{ \para \in \mathbb{R}^P: \mathcal{R}(\para) \in \mathcal{T} \}$ for some interval $\mathcal{T} \subseteq \mathbb{R} \cup \{ \infty, - \infty \}$. If we target the exceedance probability $\mathbb{P}(\mathcal{R}(\para) \geq T)$ for some real number $T$, we assign $\mathcal{T} = [T, \infty)$. We are interested in $\mathbb{P}(\para \in A | \data)$ for $\para$ distributed according to the posterior PDF $p(\para|\data)$ and write,
\begin{equation}
    \mathbb{P}(\para \in A | \data) = \int_A p(\para|\data) d \para.
\end{equation}
% Here, $\mathbb{P}(\para \in A | \data)$ is used as a slightly informal yet common notation for $\mathbb{P}(\para \in A | \dataRV = \data)$.

\subsection{Bayesian inversion and Metropolis--Hastings}
In Bayes' theorem, the posterior PDF is given by, 
\begin{align}
	p(\para | \data ) = \frac{p(\para) p(\data | \para) }{p(\data)},
\end{align}
with the prior PDF~$p(\para)$ of the model parameters, the likelihood function~$p(\data | \para)$ and the evidence~$p(\data)$. As in practice, it is often not possible to sample directly from the posterior when the forward solver $\para \mapsto \mathcal{G}(\para)$ is non-linear, sampling methods such as MCMC and SMC can be applied. \par

The most used MCMC method is the Metropolis--Hastings algorithm (MH algorithm; \citeauthor{metropolis1953equation} \citeyear{metropolis1953equation}; \citeauthor{hastings1970monte}
\citeyear{hastings1970monte}). The MH algorithm is an iterative algorithm that, in each iteration, proposes a new set of model parameter values, which is then accepted or rejected based on the acceptance probability. The choice of the proposal density is crucial, as it has to balance the trade-off between exploration and exploitation. While standard Gaussian model proposals can be applied for a model space with reduced dimension, more high-dimensional parameter spaces present considerable challenges (e.g., \citeauthor{robert2018} \citeyear{robert2018}). To ensure robustness against different discretization choices and to maintain a reasonable step size while inferring thousands of unknowns, we rely on preconditioned Crank-Nicolson proposals that preserve the prior PDF (pCN; e.g. \citeauthor{cotter2013} \citeyear{cotter2013}). The utilization of such prior-preserving proposals results in the acceptance probability being solely dependent on the likelihood values. In the field of geophysics, MCMC algorithms with model proposals that preserve the prior are known as extended Metropolis \citep{mosegaard}. The pCN proposals have been utilized for instance in a parallel tempering approach by \citet{xu2020preconditioned}.

\subsection{From Sequential Monte Carlo to \ssmc}
In this Section, we first introduce Sequential Monte Carlo for posterior inference (Section \ref{SMC_SMC_post}) and Sequential Monte Carlo for rare event estimation (Section \ref{SMC_SMC_rare}). Subsequently, we introduce \ssmc, a novel sequential combination of both methods, designed to tackle the challenge of estimating rare event probabilities while accounting for posterior uncertainty (Section \ref{SMC_dsmc}). For the methodology of the first stage (Section \ref{SMC_SMC_post}), we rely on the framework of \citet{del2007sequential} and \citet{zhou2016toward} and refer to their works for further details such as convergence behaviour. Likewise, for the second stage (Section \ref{SMC_SMC_rare}), we follow the framework presented by \citet{cerou2012sequential} and suggest consulting their paper for additional information.

\subsubsection{Sequential Monte Carlo for posterior inference}
\label{SMC_SMC_post}
Posterior estimation with the SMC method is based on a particle approximation using $N$ particles $\{\para^{(1)}, \para^{(2)},...,\para^{(N)}\}$ with weights $\{W^{(1)}, W^{(2)},...,W^{(N)}\}$. If the particles are sampled according to the posterior, the weights are redundant and reduce to $1/N$. In practice, it is generally not possible to sample from the posterior and importance sampling using a density $\eta(\para | \data)$ is applied. Importance sampling generates samples from an importance distribution that assigns higher probabilities to regions where the target distribution is expected to have most of its mass, thereby reducing the variance of estimators (e.g. \citeauthor{owen2000safe} \citeyear{owen2000safe}). To achieve a well-working importance sampling approach for the posterior PDF, one should strive for a $\eta(\para|\data)$ as close as possible to $p(\para|\data)$. This can be achieved by building a sequence of $K$ PDFs $\{p_0(\para|\data), p_1(\para|\data),..., p_K(\para|\data)\}$ with $p_0(\para|\data) = p(\para)$ and $p_K(\para|\data) = p(\para|\data)$, thus moving gradually from the prior PDF to the posterior PDF (\citeauthor{del2007sequential} \citeyear{del2007sequential}). The sequence is built on unnormalized power posteriors (\citeauthor{neal2001annealed} \citeyear{neal2001annealed}),
\begin{equation}
    \label{SMC_temp}
    p_k(\para|\data) = p(\data|\para)^{\alpha_k}p(\para),
\end{equation}
with $0 = \alpha_0 < \alpha_1 < ... < \alpha_K = 1$. With increasing exponent $\alpha_k$, the relative influence of the likelihood on the power posterior grows. For a smaller exponent, the exponeniated term is `flatter' such that the power posterior is closer to the prior PDF. When using the importance density $\eta(\para|\data)$ to sample the particles $\para^{(p)}$, the weights $W^{(p)}$ correspond to the normalized version of the importance weights $w^{(p)} = p(\para^{(p)}|\data)/\eta(\para^{(p)}|\data)$.   \par

\begin{figure}	
		\centering
		\includegraphics[width=\linewidth]{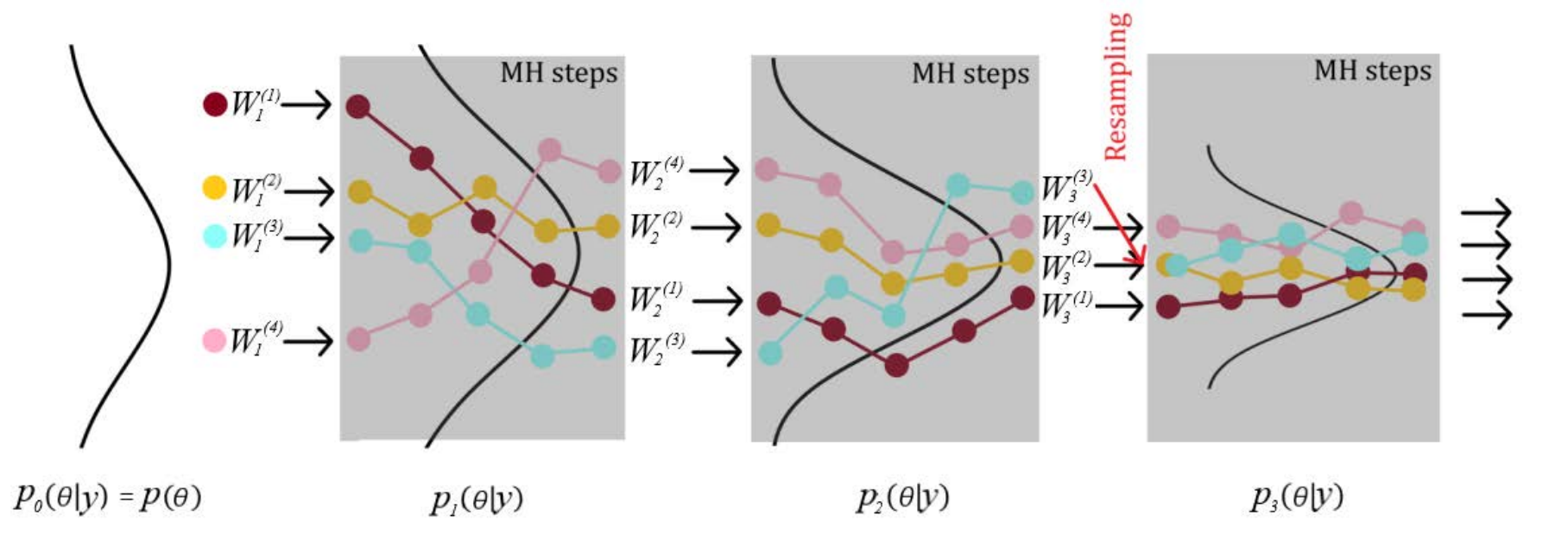}
		\caption{Illustration of the SMC method for posterior inference. We depict the first four power posteriors for an example with $N=4$ particles and $s_P=4$ MH steps.}
		\label{SMC_fig:illu_post}		
\end{figure}

We start at iteration $k=0$ with particles $\para^{(p)}_0$ ($p=1,2,...,N$) sampled from the prior PDF $p_0(\para|\data) = p(\para)$ and initial weights $W^{(p)}_0$ being all equal to $1/N$ . At iteration $k$ of the SMC method, $p_k(\para|\data)$ is approximated by importance sampling based on the previously estimated power posterior $p_{k-1}(\para|\data)$. Therefore, the particles $\para_{k-1}^{(p)}$ are assigned with incremental weights, 
\begin{equation}
    \label{SMC_SMC_weights}
    w_{k}^{(p)} =  \frac{p_{k}\left(\para_{k-1}^{(p)}|\data \right)}{p_{k-1}\left(\para_{k-1}^{(p)}|\data \right)} = \frac{p \left(\data|\para_{k-1}^{(p)} \right)^{\alpha_k}}{p \left(\data|\para_{k-1}^{(p)} \right)^{\alpha_{k-1}}} = p \left(\data|\para_{k-1}^{(p)} \right)^{\alpha_k - \alpha_{k-1}}.
\end{equation}
To account for the previous importance sampling steps, the cumulative normalized weights $W_k^{(p)}$ of the particles $\para_{k-1}^{(p)}$ are defined as,
\begin{equation}
\label{SMC_SMC_nweights}
    W_k^{(p)} =  \frac{W_{k-1}^{(p)} w_{k}^{(p)}}{\sum_{j=1}^{N} W_{k-1}^{(j)} w_{k}^{(j)}}, 
\end{equation}
taking into account the history of weights and normalizing them to ensure their sum equals one. The particles $\para_{k-1}^{(p)}$ approximating $p_{k-1}(\para|\data)$ are generated by propagating each particle $\para_{k-2}^{(p)}$ according to a Markov kernel leaving $p_{k-1}(\para|\data)$ invariant (\citeauthor{neal2001annealed} \citeyear{neal2001annealed}). This can be achieved by employing a finite number $s_P$ of MH steps \citep{del2007sequential}. In contrast to MCMC methods, the MH steps used within the SMC method do not have to converge as the importance sampling weights account for any possible sampling from the wrong distribution (\citeauthor{del2007sequential} \citeyear{del2007sequential}). The SMC procedure for posterior inference is illustrated in Figure~\ref{SMC_fig:illu_post}. \par

When the (empirical) variance of the weights $W_k^{(p)}$ at iteration $k$ becomes large, it is beneficial to resample the particles before propagation (\citeauthor{del2007sequential} \citeyear{del2007sequential}, \citeauthor{doucet2009tutorial} \citeyear{doucet2009tutorial}). Resampling decreases the variance of the weights by discarding most particles with low weights and preferably reproducing those with high weights. Here, we use systematic resampling (\citeauthor{doucet2009tutorial} \citeyear{doucet2009tutorial}). Subsequently, the weights $W_k^{(p)}$ are set to $1/N$, as the resampled particles are approximately distributed according to $p_k(\para|\data)$. Resampling increases the variance of the estimator, making it wasteful if the importance weights do not exhibit significant variability \citep{del2006sequential}. To decide when resampling is to be performed, the effective sample size (ESS; \citeauthor{kong1994sequential} \citeyear{kong1994sequential}),
\begin{equation}
\label{SMC_ESS}
    ESS_k = \frac{ \left(\sum_{p=1}^N W_{k-1}^{(p)} w_k^{(p)} \right)^2}{\sum_{o=1}^N \left(W_{k-1}^{(p)} \right)^2 \left(w_k^{(p)} \right)^2},
\end{equation}
is used. For instance, \citet{del2006sequential} apply the decision rule of resampling if the $ESS_k$ falls below 30 $\%$ of the number of particles $N$. To ensure that the final particles are a (unweighted) approximation of the posterior, we enforce a resampling step in the last iteration.  \par

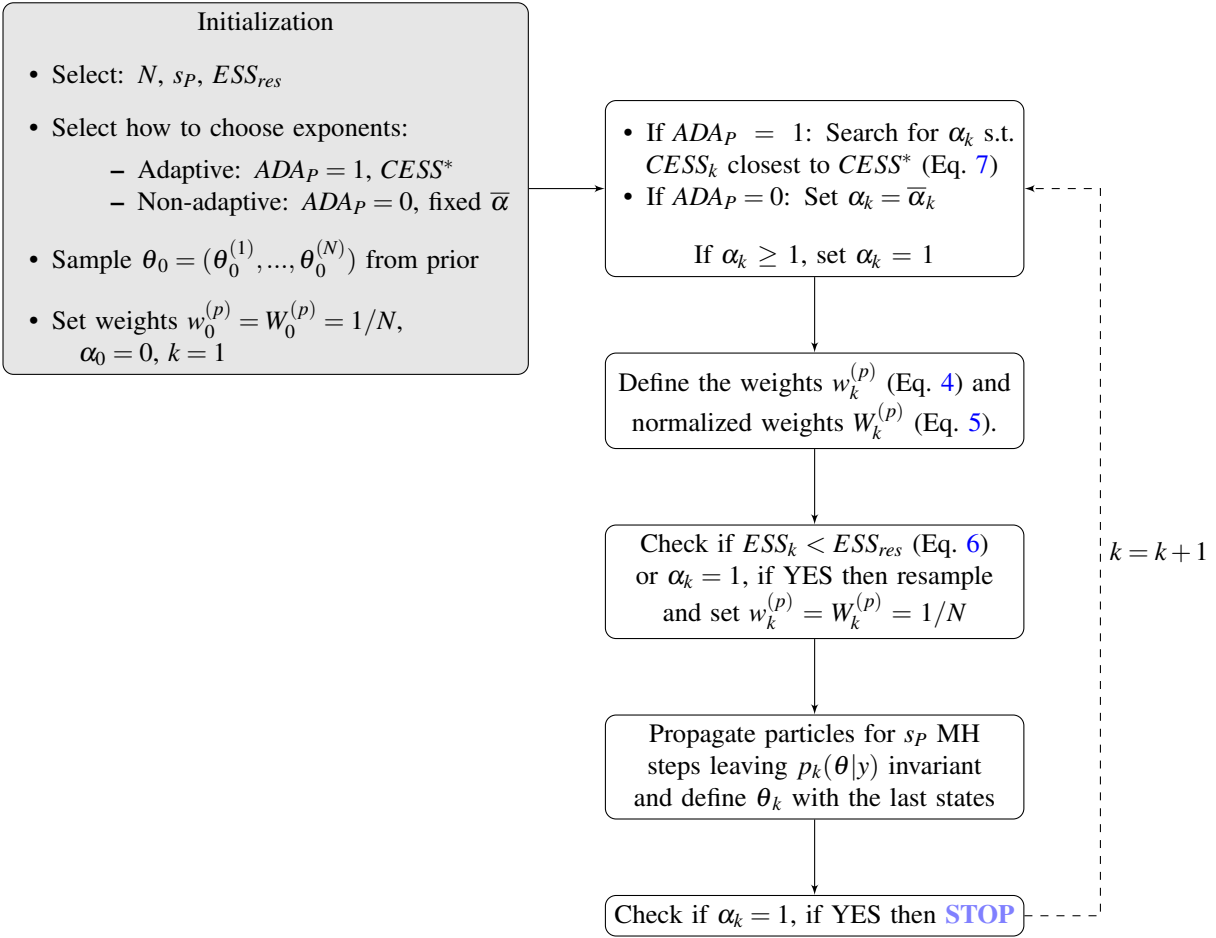
\begin{figure}[h]
	\centering
	% Define block styles
    \tikzstyle{block} = [rectangle, draw, fill=black!10, text width=19em, text centered, rounded corners, node distance=4cm]
    \tikzstyle{line} = [draw, -latex']
	\begin{tikzpicture}[minimum size=5mm,
    node distance=4cm and 7cm,
    >=stealth,
    bend angle=45,
    auto]
    % Place nodes
    \node [block] (zero) {Initialization 
    \begin{itemize}
    \setlength{\itemindent}{-1em}
        \item Select: $N$, $s_P$, $ESS_{res}$
        \item Select how to choose exponents:\\
        \begin{itemize}[nosep]
            \item Adaptive: $ADA_{P} = 1$, $CESS^*$
            \item Non-adaptive: $ADA_{P}=0$, fixed $\overline{\alpha}$
        \end{itemize}
        \item Sample $\para_0=(\para_0^{(1)},...,\para_0^{(N)})$ from prior 
        \item Set weights $w_0^{(p)} = W_0^{(p)} = 1/N$, \\ $\alpha_0 = 0$, $k=1$
         \end{itemize}};
    \tikzstyle{block} = [rectangle, draw, fill=white, text width=15em, text centered, rounded corners, node distance=3cm]
    \node [block, right=1cm of zero] (one) {
    \begin{itemize}[noitemsep,topsep=0pt, leftmargin=12pt]
        \item If $ADA_{P}=1$: Search for $\alpha_k$ s.t. $CESS_k$ closest to $CESS^*$ (Eq.~\ref{SMC_CESS}) 
        \item If $ADA_{P}=0$: Set $\alpha_k = \overline{\alpha}_k$
    \end{itemize}
    \flushleft{If $\alpha_k \geq 1$, set $\alpha_k = 1$}
    };
    \node [block, below=1cm of one] (two) {Define the weights $w^{(p)}_k$ (Eq.~\ref{SMC_SMC_weights}) and normalized weights $W^{(p)}_k$ (Eq.~\ref{SMC_SMC_nweights}).};
    \node [block, below=1cm of two] (three) {Check if $ESS_k < ESS_{res}$ (Eq.~\ref{SMC_ESS}) or $\alpha_k = 1$, if YES then resample and set $w_k^{(p)} = W_k^{(p)}=1/N$ \\
    };
    \node [block, below=1cm of three] (four) {Propagate particles for $s_P$ MH steps leaving $p_k(\para|\data)$ invariant and define $\para_k$ with the last states    };
    \node [block, below=1cm of four] (five) {Check if $\alpha_k = 1$, if YES then \textcolor{blue!50}{\textbf{STOP}}};
    \path [line] (zero) -- (one) ;
    \path [line] (one) -- (two);
    \path [line] (two) -- (three);
    \path [line] (three) -- (four);
    \path [line] (four) -- (five);
    \path [dashed, line] (five.east) -- ++(1,0)  |- (one)  node[near start,anchor=west] {$k=k+1$ } {};
\end{tikzpicture}
	\caption{Flow chart illustrating the SMC method for posterior inference.}
	\label{SMC_fig:flow_post}
\end{figure}

When defining the sequence of exponents $\alpha$, one has to consider that too large differences between $\alpha_{k-1}$ and $\alpha_k$ lead to a large discrepancy between the power posteriors $p_{k-1}(\para|\data)$ and $p_k(\para|\data)$ and a subsequent high variance of the importance sampling estimator. However, if the difference is very small, an excessive number of steps are needed until $\alpha_k=1$ is reached. It is natural to aim for a similar discrepancy between successive power posteriors (\citeauthor{zhou2016toward} \citeyear{zhou2016toward}). To select the sequence of exponents $\alpha$, we use the adaptive method of \citet{zhou2016toward}, based on the conditional effective sample size (CESS), 
\begin{equation}
\label{SMC_CESS}
    CESS_k = N \frac{ \left(\sum_{p=1}^N W_{k-1}^{(p)} w_k^{(p)} \right)^2}{\sum_{p=1}^N W_{k-1}^{(p)} \left(w_k^{(p)} \right)^2}.
\end{equation}
The $CESS_k$ quantifies the quality of $p_{k-1}(\para|\data)$ as an importance density to estimate expectations under $p_{k}(\para|\data)$ (\citeauthor{zhou2016toward} \citeyear{zhou2016toward}). The $CESS$ is equal to the $ESS$ when resampling is conducted at each iteration. \citet{zhou2016toward} show that using the $CESS$ for the adaptive sequence leads to a reduction in estimator variance compared to an approach using the $ESS$. To define the next $\alpha_k$, a binary search for the value for which the $CESS$ is the closest to a pre-defined target value $CESS^{*}$ is performed \citep{zhou2016toward}. A binary search operates by iteratively halving the interval containing the potential values, effectively reducing the search space with each step by comparing the target value to the middle element. If the target is less than the middle element, the search is restricted to the lower half of the interval; if it's greater, the search is limited to the upper half. The closer this target value $CESS^{*}$ is to $N$, the better the approximation, but the slower the algorithm becomes as the number of power posteriors grows. The SMC algorithm stops when $\alpha_k$ reaches one. Such an adaptive approach is expected to result in a more efficient algorithm compared to its non-adaptive counterpart. Importantly, it also leads to a more automated algorithm by minimizing the number of user-defined tuning parameters \citep{beskos2016convergence}. However, using an adaptive method for the selection of the exponents introduces a slight bias into the results. \citet{beskos2016convergence} explore the convergence behaviour for such adaptive approaches and establish that the output satisfies a weak law of large numbers and a central limit theorem. To indicate if we use an adaptive or fixed sequence of exponents, we specify the binary variable $ADA_P$ as 1 for an adaptive and 0 for a predetermined selection. The full workflow of the SMC method for posterior inference is summarized in Figure \ref{SMC_fig:flow_post}. 

\subsubsection{Sequential Monte Carlo for rare event estimation}
\label{SMC_SMC_rare}
The SMC method can be modified to enable simulation of rare events and estimation of their probabilities by using a sequence of not-so-rare nested events \citep{del2006sequential, johansen2005sequential, cerou2012sequential}. It is assumed that $\para$ is a random element on $\mathbb{R}^P$ with probability distribution $p(\para)$ that can be sampled from. To estimate $\mathbb{P}(\para \in A)$, the SMC method for rare event estimation employs a sequence of nested sets $A_k = \{ \para \in \mathbb{R}^P: \mathcal{R}(\para) \in \mathcal{T}_k \}$, with $\mathbb{R}^P = A_0 \supset A_1 \supset ... \supset A_K = A$. It holds that, 
\begin{equation}
\label{SMC_risk_esti_form}
    \mathbb{P} \left(\para \in A \right) = \prod\limits_{k=1}^{K} \mathbb{P} \left(\para \in A_{k} | \para \in A_{k-1} \right).
\end{equation}
If we are interested in $\mathbb{P}(\mathcal{R}(\para) \geq T)$, the sequence of nested sets $A_k = \{ \para \in \mathbb{R}^P: \mathcal{R}(\para) \in [T_k, \infty) \}$ corresponds to a sequence of increasing thresholds $\{ T_0, ...,T_K \}$ with $T_0=-\infty$ and $T_K = T$. For $\mathbb{P}(\mathcal{R}(\para) \leq T)$, we employ $A_k = \{ \para \in \mathbb{R}^P: \mathcal{R}(\para) \in (-\infty,T_k] \}$ using a sequence of decreasing thresholds with $T_0=\infty$ and $T_K = T$. \par

\begin{figure}	
		\centering
		\includegraphics[width=\linewidth]{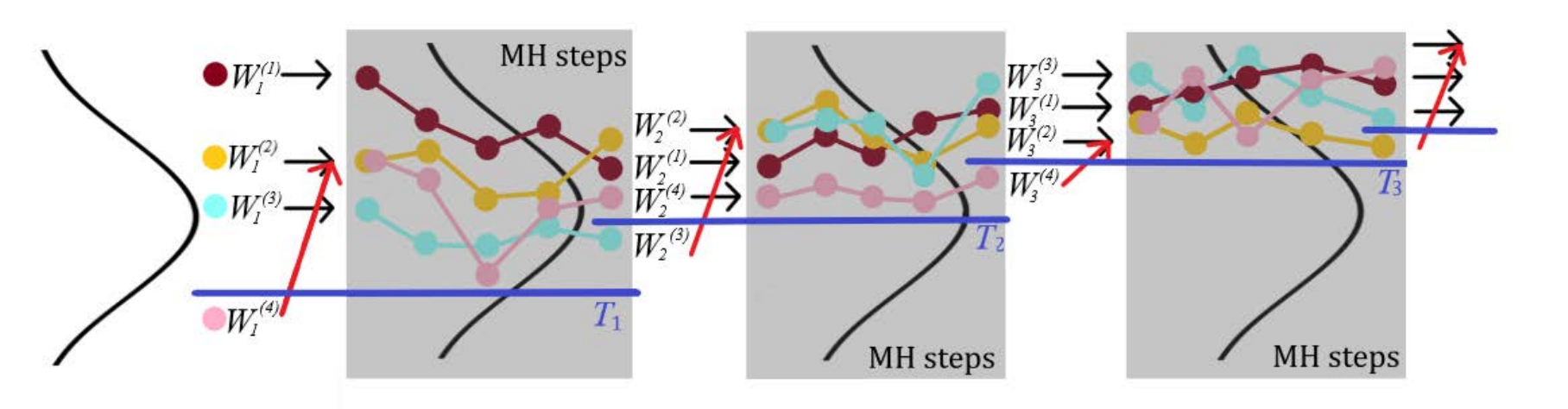}
		\caption{Illustration of the SMC method for rare events targeting $\mathbb{P}(\mathcal{R}(\para) \geq T)$. We depict the first three thresholds for an example with $N=4$ particles, $s_R=4$ MH steps and a quantile of $\gamma=0.25$.}
		\label{SMC_fig:illu_rare}				
\end{figure}

The SMC method for rare event estimation starts by initializing $N$ particles $\para_0 = (\para_0^{(1)},...,\para_0^{(N)})$ sampled from $p(\para)$. The first intermediate distribution $p_{A_0}(\para) = p(\para | \para \in A_{0})$ is equal to $p(\para)$. To approximate the intermediate distribution $p_{A_k}(\para) = p(\para | \para \in A_{k})$ for $k \geq 1$, each particle $\para_{k-1}^{(p)}$ is assigned a weight, 
\begin{equation}
    \label{SMC_SMC_nweights_risk}
    W_{k}^{(p)} = \begin{cases}
    1/|I_k|, \text{ if } \para_{k-1}^{(p)} \in A_k\\
    0, \text{ otherwise, }
    \end{cases}
\end{equation}
with $I_k = \{p: \para_{k-1}^{(p)} \in A_k \}$ and $|I_k|$ denoting its cardinality. Thereby, we are assuming that $I_k$ is non-empty, otherwise the particle system dies. Subsequently, systematic resampling (\citeauthor{doucet2009tutorial} \citeyear{doucet2009tutorial}) is employed such that particles which do not lie in $A_k$ are replaced by particles that do. The resampled particles are propagated using a Markov kernel, leaving $p_{A_k}(\para)$ invariant (\citeauthor{cerou2012sequential} \citeyear{cerou2012sequential}). We are considering $s_R$ steps with a MH algorithm whereby a transition is only accepted if $\para$ stays in $A_k$. The procedure of SMC for rare event estimation targeting $\mathbb{P}(\mathcal{R}(\para) \geq T)$ is illustrated in Figure~\ref{SMC_fig:illu_rare}.	\par

\begin{figure}[h]
	\centering
	% Define block styles
    \tikzstyle{block} = [rectangle, draw, fill=black!10, text width=19em, text centered, rounded corners, node distance=4cm]
    \tikzstyle{line} = [draw, -latex']
	\begin{tikzpicture}[minimum size=5mm,
    node distance=4cm and 7cm,
    >=stealth,
    bend angle=45,
    auto]
    % Place nodes
    \node [block] (zero) {Initialization 
    \begin{itemize}
    \setlength{\itemindent}{-1em}
        \item Select: $N$, $s_R$ 
        \item Select how to choose thresholds:\\
        \begin{itemize}[nosep]
            \item Adaptive: $ADA_{R} = 1$, $\gamma$-quantile
            \item Non-adaptive: $ADA_{R}=0$, fixed $\overline{T}$
        \end{itemize}
        \item Sample $\para_0 = (\para_0^{(1)},...,\para_0^{(N)})$ from prior
        \item Set weights $W_0^{(p)} = 1/N$,  \\
        $\mathbf{T} = (T_0, T_1,...) = (-\infty, -\infty,...)$, \\
        $P = (P_1, P_2,...) = (1, 1,...) $ 
         \end{itemize}};
    \tikzstyle{block} = [rectangle, draw, fill=white, text width=15em, text centered, rounded corners, node distance=3cm]
    \node [block, right=1cm of zero] (one) {
    \begin{itemize}[noitemsep,topsep=0pt, leftmargin=12pt]
        \item If $ADA_{R}=1$: Next threshold \\ with $\gamma$-quantile (Eq. \ref{SMC_thres})
        \item If $ADA_{R}=0$: Set $T_k = \overline{T}_k$
    \end{itemize}
    \flushleft{ Check if $T_k \geq T$}
    };
    \node [block, below=1cm of one] (two) {Define the weights $W_k^{(p)}$ (Eq.~\ref{SMC_SMC_nweights_risk}).};
    \node [block, below=1cm of two] (three) {Resample particles according to weights and set $P_k = |I_k|/N$ (Eq.~\ref{SMC_SMC_nweights_risk})
    };
    \node [block, below=1cm of three] (four) {Propagate particles for $s_R$ MH steps leaving $p_{A_k}(\para)$ invariant and define $\para_k$ with the last states. };
    \tikzstyle{block} = [rectangle, draw, fill=white, text width=12em, text centered, rounded corners, node distance=3cm]
    \node [block,above right=1cm and -3cm  of one, fill=blue!50, fill opacity=0.3, text opacity=1] (two3) {Set $T_k = T$, \\ $P_k = |I_k|/N$ and calculate 
    $\mathbb{P} (\para \in A ) = \prod_{j=1}^{k} P_j$};
    \path [line] (zero) -- (one) ;
    \path [line] (one) -- node[anchor=east] {NO} (two);
    \path [line] (two) -- (three);
    \path [line] (three) -- (four);
    \path [dashed, line] (four.east) -- ++(1,0)  |- (one)  node[near start,anchor=west] {$k=k+1$ } {};
    \path [line, color=blue!50] (one) -- node[anchor=east] {YES } (two3);
\end{tikzpicture}
	\caption{Flow chart illustrating the SMC method for rare event estimation of $\mathbb{P}(\mathcal{R}(\para) \geq T)$.}
	\label{SMC_fig:flow_rare}
\end{figure}
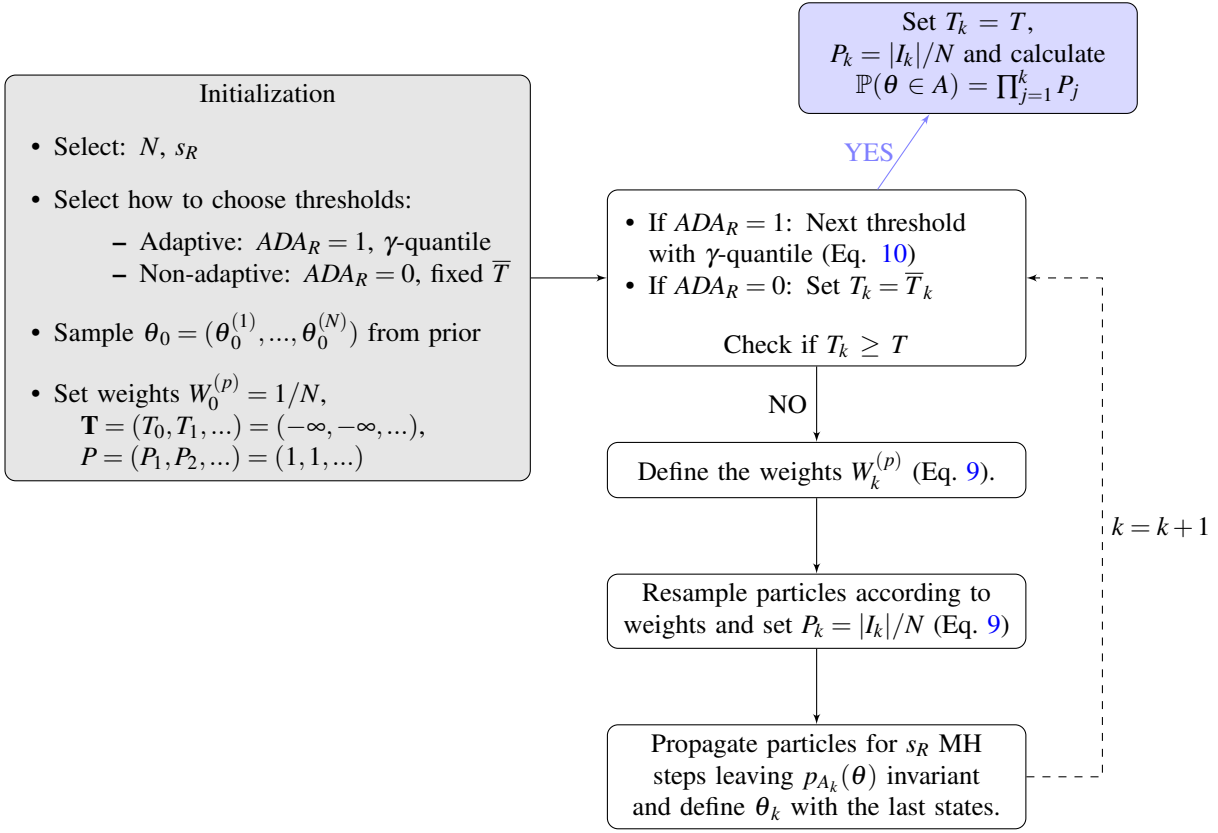

We need to choose a sequence of nested sets such that $\mathbb{P} (\para \in A_{k} | \para \in A_{k-1})$ is reasonably high.  \citet{cerou2012sequential} detail both a fixed and an adaptive algorithm. For $A_k = \{ \para \in \mathbb{R}^P: \mathcal{R}(\para) \in [T_k, \infty) \}$, an adaptive method based on quantiles of $\mathcal{R}(\cdot)$ of the particles ensures that the asymptotic variance of the estimator is minimal (see \citeauthor{cerou2012sequential} \citeyear{cerou2012sequential}). Utilizing the $\gamma$-quantile,
\begin{equation}
\label{SMC_thres}
    T_k = q_{\gamma} \left( \mathcal{R}(\para_{k-1}) \right) ,
\end{equation}
guarantees that a ratio of $(1-\gamma)$ of the particles survive. The adaptive algorithm's stopping criterion is met when the quantile surpasses the targeted threshold, at which point the last $T_K$ is set equal to $T$. Then, $\mathbb{P}(\para \in A)$ is estimated by multiplication of all $P_k = |I_k|/N$ for $k=1,...,K$. Due to the adaptiveness of the thresholds, the resulting estimator is biased given the finite number of particles $N$ \citep{au2001estimation}. This bias is positive and becomes negligible compared to the variance of the estimator as the number of particles increases \citep{cerou2012sequential}. To circumvent this bias, one can either re-run the algorithm with the previously optimized sequence or use a predetermined fixed sequence of thresholds. With the binary variable $ADA_R$ we indicate if we use fixed ($ADA_R=0$) or adaptive ($ADA_R=1$) sequences of thresholds. The work flow of the SMC method for rare event estimation is summarized in the flow chart in Figure \ref{SMC_fig:flow_rare}. \par

\subsubsection{Posterior Risk Sequential Monte Carlo method}
\label{SMC_dsmc}
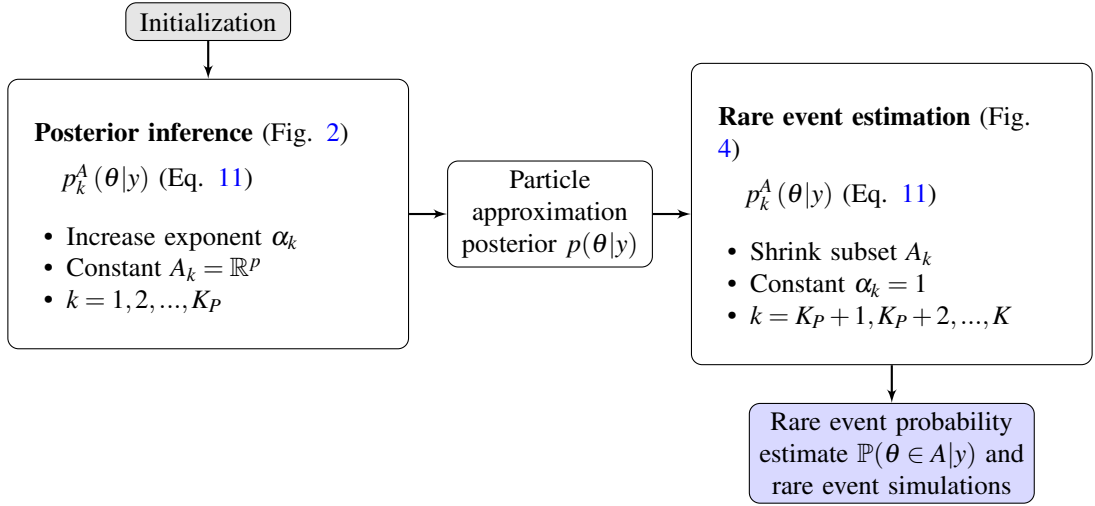
\begin{figure}
	\centering
	% Define block styles
    \tikzstyle{block} = [rectangle, draw, fill=black!10, text width=5.5em, text centered, rounded corners, node distance=4cm]
    \tikzstyle{line} = [draw, -latex']
	\begin{tikzpicture}[minimum size=5mm,
    node distance=4cm and 7cm,
    >=stealth,
    bend angle=45,
    auto]
    % Place nodes
    \node [block] (zero) {Initialization };
    \tikzstyle{block} = [rectangle, draw, fill=white, text width=13em, text centered, rounded corners, node distance=3cm, inner xsep=1em]
    \node [block, below=0.5cm of zero] (one) {
    \flushleft{\textbf{Posterior inference} (Fig. \ref{SMC_fig:flow_post}) }\\[5pt]
    $\quad p_k^A \left(\para|\data \right)$ (Eq. \ref{SMC_powerriskpost}) \\[5pt]
    \begin{itemize}[noitemsep,leftmargin=12pt]
        \item Increase exponent $\alpha_k$
        \item Constant $A_k = \mathbb{R}^{p}$
        \item $k=1,2,...,K_P$
        \item[]
        % \item $s_P$ MH steps leaving $p_k^A$ invariant
    \end{itemize}};
    \tikzstyle{block} = [rectangle, draw, fill=white, text width=7em, text centered, rounded corners, node distance=3cm]
    \node [block, right=.5cm of one] (one1) 
    { Particle approximation posterior $p(\para|\data)$
    };
    \tikzstyle{block} = [rectangle, draw, fill=white, text width=13em, text centered, rounded corners, node distance=3cm, inner xsep=1em]
    \node [block, right=.5cm of one1] (two) {
    \flushleft{\textbf{Rare event estimation} (Fig. \ref{SMC_fig:flow_rare}) }\\[5pt]
    $\quad p_k^A \left(\para|\data \right)$ (Eq. \ref{SMC_powerriskpost})\\[5pt]
    \begin{itemize}[noitemsep,leftmargin=12pt]
        \item Shrink subset $A_k$ 
        \item Constant $\alpha_k=1$ 
        \item $k=K_P+1,K_P+2,...,K$
        \item[]
        % \item $s_R$ MH steps leaving $p_k^A$ invariant
    \end{itemize}};
    \tikzstyle{block} = [rectangle, draw, fill=white, text width=10em, text centered, rounded corners, node distance=3cm]
    \node [block, below=.5cm of two, fill=blue!50, fill opacity=0.3, text opacity=1] (three) 
    {Rare event probability estimate $\mathbb{P}(\para \in A| \data)$ and rare event simulations
    };
    \path [line, thick] (zero) -- (one) ;
    \path [line, thick] (one) -- (one1);
    \path [line, thick] (one1) -- (two);
    \path [line, thick] (two) -- (three);
\end{tikzpicture}
	\caption{Work flow of the \ssmc method.}
	\label{SMC_fig:flow1}
\end{figure}

To estimate $\mathbb{P}(\para \in A| \data)$, we introduce a sequential combination of the two SMC methods described in Sections \ref{SMC_SMC_post} and \ref{SMC_SMC_rare} \ssmcbracket. Let us write the $k$-th power posterior with respect to the subset $A_k$ as,
\begin{equation}
\label{SMC_powerriskpost}
 p_k^A \left(\para|\data \right) = p \left(\data|\para \right)^{\alpha_k}p(\para) \mathbbm{1}{\{ \para \in A_k \}}.   
\end{equation}
While the first stage of the \ssmc algorithm generates particles distributed according to the posterior by increasing the exponent of the likelihood $\alpha_k$ with the subset $A_k$ being held constant as $\mathbb{R}^p$, the second stage shrinks the subset while leaving the exponent of the power posterior at $1$. For the rare event analysis, it is crucial that we start the second phase with a unweighted particle approximation of the posterior, ensured by the resampling step in the last step of the posterior inference stage. We denote as $K_P$ the number of intermediate power posteriors, as $K_R$ the number of thresholds and as $K = K_P + K_R$ their sum. Additionally, we define $s_P$ as the number of MH steps employed between each importance sampling step in the posterior phase and $s_R$ as the number between the subset sampling steps during the rare event phase. When the same number of steps is used for both, we denote it as $s = s_P = s_R$. The \ssmc method inherits the theoretical properties of the SMC methods utilized in the two stages, including any biases present in the estimators resulting from adaptive sequences of exponents and thresholds. The complete work flow of the \ssmc method is summarized in Figure~\ref{SMC_fig:flow1}. \par

In high-dimensional scenarios characterized by complex posterior distributions, the process of particle propagation using a limited number of MH steps can become limiting. In such contexts, the frequency of particle resampling becomes important to monitor. In the rare event probability estimation phase, this aspect becomes even more critical as frequent resampling is unavoidable. This implies the need to ensure that a sufficient number of MH steps are used to prevent particle collapse following the resampling steps. \par

In groundwater settings where the rare event revolves around contamination hazards, the simulation of the quantity of interest often demands more computational resources than the forward model used to estimate the posterior PDF. To achieve computational speed-up under such situations (as exemplified in Section \ref{SMC_ssmc_2d}), we introduce a minor modification to the propagation step during the rare event phase of \ssmc. Instead of simulating both the forward response and quantity of interest in every step, we conduct first a series of $ss_R$ posterior steps within each of the $s_R$ steps. Subsequently, the last state is treated as a proposal from the posterior which is accepted or rejected based on whether it falls within the current subset.

\section{1D flow example}
\label{SMC_ssmc_1d}
As a first example, we study a steady-state 1-D groundwater flow problem (diffusion equation). The chosen problem setting is inspired by a test case by \citet{straub2016bayesian}, which corresponds to the steady-state version of a test case introduced by \citet{marzouk2009dimensionality}. The fast run time of this simple toy example allows for a sensitivity analysis of the algorithmic parameters of the \ssmc method.

\subsection{Synthetic setting} % units m$^2$/s; 
\label{SMC_synth_1D}
The model domain is the unit interval $D=[0,1]$ m and we consider the following steady-state equation,
\begin{equation}
\label{SMC_diff_eq}
    \frac{d}{dx} \left( \para(x)\frac{dh}{dx} \right) + b(x) = 0,
\end{equation}
with hydraulic conductivity $\para(x)$ [m/s], source $b(x)$ [1/s] and hydraulic head $h(x)$ [m]. \par

\begin{figure}	
		\centering
        \includegraphics[width=\textwidth]{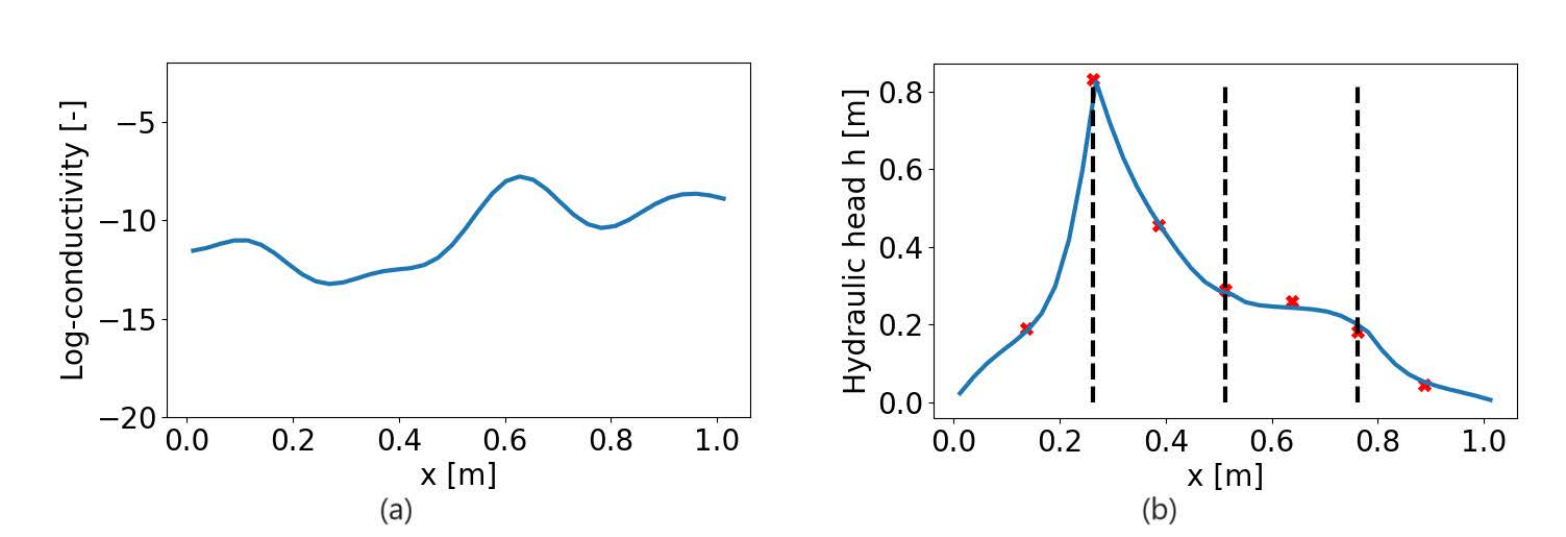}
        \caption{(a) 'True' log-hydraulic conductivity $\log \para(x)$ on $D=[0,1]$ m and corresponding (b) steady-state solution $h(x)$ (solid line) for the diffusion equation including the pumping sources (source locations dashed) and the resulting noisy measurements $\data$ (crosses).}
		\label{SMC_fig:log_difff}				
\end{figure}

The log-conductivity $\log \para(x)$ is parameterized as a finite rank Gaussian random process expressed by,
\begin{equation}
\label{SMC_karhunen}
    \log \para(x) = \mu_{\log \para} + \sum_{i=1}^{n} \sqrt{w_i} v_i(x) Z_i,
\end{equation}
with $\{w_i, v_i\}$ representing the first $n$ eigenvalues and eigenfunctions from the Karhunen-Loève expansion \citep{loeve1977elementary} of a Gaussian process with mean $\mu_{\log \para} = \log(10^{-5})$ and exponential covariance function $\kappa_{\log \para}(\Delta x) = \sigma^2 \exp(-\Delta x/l)$ with standard deviation $\sigma = 3$ and integral scale $l=0.3$ m. $Z_i$ denote independent standard normally-distributed variables. Following \citet{straub2016bayesian}, we employ a truncation after $n=10$ terms. For the representation, we use a uniform grid with 40 intervals and under the assumption of the mean and covariance structure being known, we infer the ten first $Z_i$. The 'true' log-hydraulic conductivity values $\log \para(x)$ are depicted in Figure~\ref{SMC_fig:log_difff}a. \par

For the measurements, the source term $b(x)$ in Equation (\ref{SMC_diff_eq}) is modelled using sources in the cells at 0.26, 0.51 and 0.76 m with identical strengths of 0.001 1/s. The measurements $\data$ are performed on the steady-state solution of $h(x)$ employing 7 sensors spaced uniformly on $D$ excluding the endpoints. To achieve this, Equation (\ref{SMC_diff_eq}) is solved with linear finite differences on a uniform grid employing 40 cells and boundary conditions $h(0)=h(1)=0$ m (\citeauthor{langtangen2017diffusion} \citeyear{langtangen2017diffusion}). Finally, the synthetically-generated measurement values are contaminated with independent Gaussian errors having a standard deviation of 0.01 m (Fig.~\ref{SMC_fig:log_difff}b). \par

For the rare event, we consider flow from the left to the right of the model domain and define the `hazard' as the flow rate on the right boundary exceeding a critical value of $T$. To calculate the flow rate, we assume a hydraulic head difference of 1 m and take the harmonic mean of the conductivity values. To enable a comparison with MC estimation, we consider a first value of $T^* = 9 \times 10^{-6}$ m/s; the second value of $T^{**}= 9.5 \times 10^{-6}$ m/s is selected such that it targets a rare event with probability of one in a billion.  \par

\subsection{Results}
We employ independent normal prior PDFs for the unknown $Z_i$ of the KL-expansion representing the log-conductivity (Eq. \ref{SMC_karhunen}). For the likelihood, we assume independent Gaussian measurement errors with the same standard deviation as used in the data generation process. We compare the results of the \ssmc method with those of a standard MH algorithm employing Gaussian proposals. To ensure an acceptance rate of approximately 30 $\%$, the step width of the proposals is adjusted accordingly, taking into account the different scales of variation in the KL components (based on initial MH runs). The same configuration of the MH algorithm is used in the MH steps employed in each iteration of the \ssmc method. \par

\begin{figure}	
		\centering
        \includegraphics[width=\linewidth]{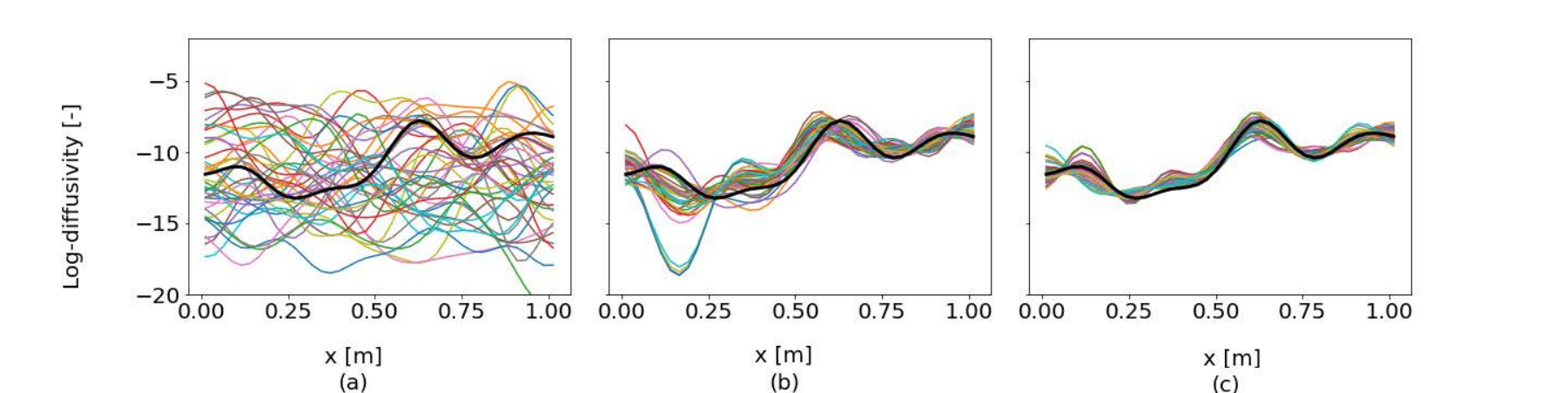}
     \caption{Results for the 1D flow example with the \ssmc method: Particle representation ($N=40$) of the log-conductivity's (a) prior, (b) posterior and (c) posterior rare event (for $T^*$) distribution; the black lines depict the true profile and the coloured lines the particles. }
		\label{SMC_fig:1D_particles}				
\end{figure}

For the \ssmc method, the following parameter choices have to be made: the number of particles $N$, the number of MH steps $s$ in each iteration (here $s=s_P=s_R$), the selection of the exponents $\alpha_k$ (Eq. \ref{SMC_CESS}), the threshold $ESS_{res}$ below which resampling is employed (Eq. \ref{SMC_ESS}) and the selection of the thresholds $T_k$ (Eq. \ref{SMC_thres}). Following \citet{del2006sequential}, we fix $ESS_{res}=0.3 \times N$ for the resampling in the initial stage of posterior inference. We start by testing a  configuration of \ssmc with $N=40, CESS^*=0.99 \times N$, $\gamma=0.05$ and $s=40$, employing adaptive schedules for the likelihood's exponents and the thresholds. Figure~\ref{SMC_fig:1D_particles} depicts resulting particle approximations of the following distributions of the log-diffusivity profile: (a) prior $p_0^A(\para|\data) = p(\para)$, (b) posterior $p_{K_P}^A(\para|\data) = p(\data|\para)p(\para)$ and (c) posterior rare event $p_K^A(\para|\data) = p(\data|\para)p(\para) \mathbbm{1}{\{ \mathcal{R}(\para) \geq T^{*} \}}$. Considering our utilization of only 40 particles, it is not particularly problematic or unexpected that the true value may deviate outside the expected range in specific areas of Figure ~\ref{SMC_fig:1D_particles}c. \par

\begin{figure}	
	\centering
         \includegraphics[width=\textwidth]{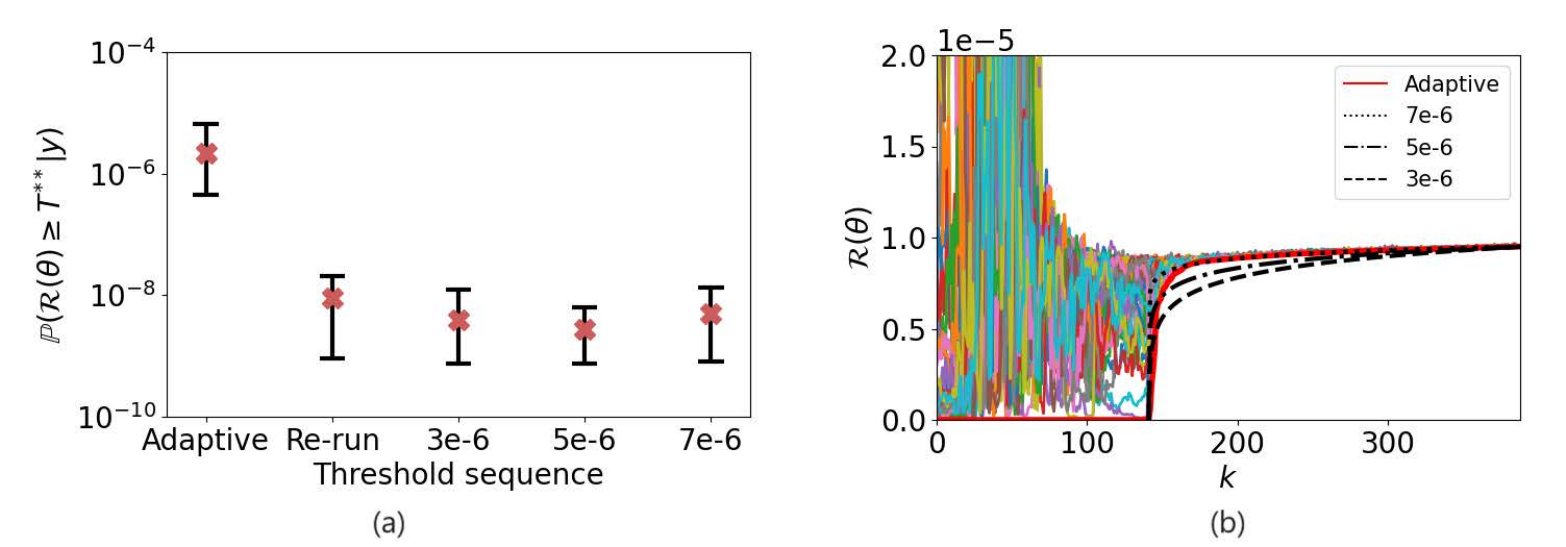}
         % \label{SMC_fig:1D_thres2}
         % \label{SMC_fig:1D_thres1}
		\caption{Illustration of the bias resulting from the adaptively determined threshold sequence within the \ssmc method for the 1D flow example: (a) Range of estimates $\mathbb{P}(\mathcal{R}(\para) \geq T^{**} | \data)$ using the different threshold sequences (ten runs each); the red crosses indicate the mean of the values and (b) evolving particle estimation of $\mathcal{R}(\para)$ with the adaptive $T_k$-sequence (red) and the different fixed logarithmic sequences (black, Eq. \ref{SMC_log_thres} with different $T_{K_P+1}$). }
		\label{SMC_fig:1D_adaptive}				
\end{figure}

To explore the level of bias introduced by the adaptive schemes for our choice of $N=40$ particles, we re-run the algorithm using the previously defined sequences as pre-defined values. The range of ten resulting estimates for $T^{**}$ are depicted in Figure~\ref{SMC_fig:1D_adaptive}a (adaptive and re-run). The adaptive runs yield a mean estimate that is approximately 200 times greater than that of the re-runs. To circumvent bias while avoiding the computational burdens associated with increasing the number of particles or performing re-runs, we adopt in what follows a fixed sequence of thresholds for the rare event estimation part ($ADA_{R}=0$ in Fig.~\ref{SMC_fig:flow_rare}). With $K_P$ denoting the number of intermediate power posteriors and following the flow chart in Figure~\ref{SMC_fig:flow1}, the first threshold different from minus infinity is $T_{K_P+1}$. For the shape of the sequence, a suitable form can be determined, for example, by conducting an initial adaptive run (Fig.~\ref{SMC_fig:1D_adaptive}b). We use a logarithmic function,
\begin{equation}
\label{SMC_log_thres}
    f_T(k) = a \log (k) + T_{K_P + 1},
\end{equation}
increasing from $T_{K_P + 1}$ to $T^{**}$. Therefore, we set the thresholds to $T_k = f_T(k-K_P)$ for $k = K_P+1,...,K$ and ensure that $T_K = f_T(K_R) = T^{**}$ by expressing $a = (T^{**}-T_{K_P + 1})/\log(K_R)$. Finally, we change the closest value of $T^{*}$ to this very value. For the first threshold, we test the choices of $T_{K_P+1}=3,5,7 \times 10^{-6}$. The resulting threshold sequences are depicted in Figure~\ref{SMC_fig:1D_adaptive}b, together with the adaptive sequence utilizing $\gamma = 0.05$. The range and mean of ten estimates for $T^{**}$ obtained with the different sequences are depicted in Figure~\ref{SMC_fig:1D_adaptive}a. We note that while the adaptive sequence leads to much higher values, the ones of the re-runs and the fixed sequences with the different $T_{K_P + 1}$ are comparable. \par

In our specific context, where the focus is on estimating the probability of rare events and the posterior of $\para$ is rather smooth, the bias caused by the adaptive schedule in the first stage of posterior estimation is minimal. Tests (not shown) demonstrated that even when considering $T^{**}$ and $N=40$, the adaptive sequence for the posterior estimation resulted in an almost identical mean estimate compared to the re-runs (less than 0.02 $\%$ difference). As a result, we continue to use an adaptive sequence of exponents for the first stage of the algorithm ($ADA_{P}=1$ in Fig.~\ref{SMC_fig:flow_post}). \par

\begin{figure}	
		\centering
        \includegraphics[width=\textwidth]{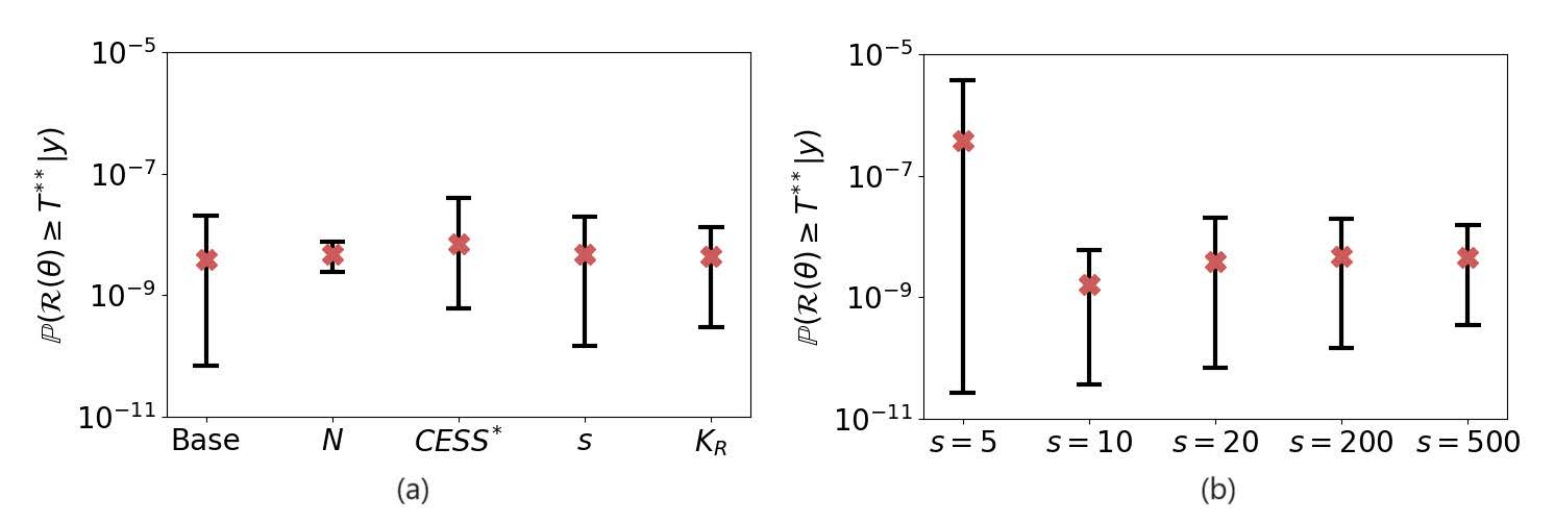}
         % \label{SMC_fig:1D_risk2_all}
         % \label{SMC_fig:1D_risk_s}
     \caption{Impact of the configuration choices within the \ssmc method for the 1D flow example. (a) Range of the rare event probability estimates for $T^{**}$ with the first bar corresponding to the base configuration and the following ones referring to the successive allocation of ten times more computational resources for either of the parameters with $N=200, CESS^*=0.9999 \times N$, $K_R = 1330$ and $s=200$. (b) Range of the rare event probability estimates for $T^{**}$ using different numbers of MH steps $s$. The red crosses in both plots indicate the mean values of the ten runs. }
		\label{SMC_fig:1D_risk_conf}				
\end{figure}

We now keep $T_{K_P+1} = 5 \times 10^{-6}$ and explore the influence of the remaining parameters on the rare event estimation. As a baseline configuration, we use $N=20, CESS^*=0.9 \times N$ (resulting in $K_P=40$), $K_R = 100$ and $s=20$, requiring 55,000 forward simulations for $T^{**}$. Next, we multiply the computational budget by a factor of ten, allocating these additional computational resources successively to each of the parameters. This results in $N=200$, $CESS^*=0.9999 \times N$ (such that $K_P=1250$), $K_R = 1330$ and $s=200$. The resulting ranges of the rare event probability estimates for $T^{**}$ using ten runs are depicted in Figure~\ref{SMC_fig:1D_risk_conf}a and the means and coefficients of variation (COV; ratio of standard deviation to the mean) for both thresholds are summarized in Table \ref{SMC_tab:diff_SMC2}. While the means are comparable for all configurations, it is seen that the parameter with the most impact in reducing the COV for both thresholds is the number of particles $N$. In this test example, the optimal $CESS^*$ only has limited influence on the variance of the rare event estimate. Still, a high-quality representation of the posterior from the first stage leads to a smaller variance of the rare event estimate. Concerning the number of MH steps, we perform additional tests with values $s=5, 10, 20, 200, 500$ (Fig.~\ref{SMC_fig:1D_risk_conf}b for $T^{**}$). While there is high variance in the estimates for $s=5$, the variance seems to stabilize from a value of $s=20$ steps. Further increasing $s$ to 200 or 500 necessitates a considerable number of additional forward operations, but leads to a much smaller improvement in the accuracy of the rare event estimate compared to increasing the number of particles. Furthermore, in the context of parallel computation, increasing the number of particles is more efficient compared to increasing $s$. Finally, when testing a value of $K_R$ smaller than 100, we observed frequent failures due to the particle system dying. On the other hand, increasing the value to $K_R=1330$ resulted in a decrease in the COV for both thresholds. Although this decrease was more significant than the effect of increasing the number of MH steps $s$, it still did not match the substantial improvement achieved by increasing the number of particles. \par

\begin{figure}	
	\centering
        \includegraphics[width=\textwidth]{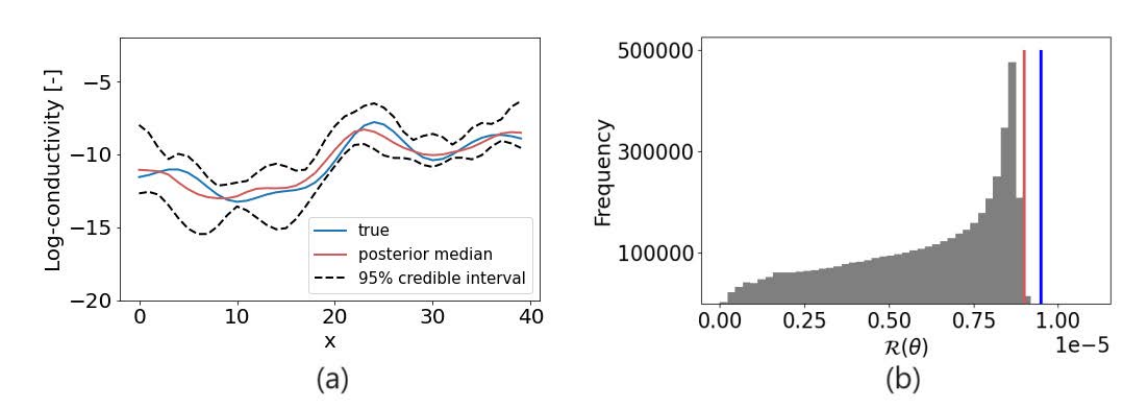}
         % \label{SMC_fig:mcmc_1_diff}
         % \label{SMC_fig:mcmc_2_diff}
		\caption{Results for the 1D flow example with the MH method: (a) Estimated posterior median (red) and credible interval (dashed) of the log-conductivity profile, together with the true profile (blue) and (b) transformed MH samples using $\para \mapsto \mathcal{R}(\para)$ with the thresholds of interest indicated ($T^*$ in red and $T^{**}$ in blue).  }
		\label{SMC_fig:mcmc_diff}				
\end{figure}

\begin{table}[]
\small
    \caption{Table summarizing the different trials of the \ssmc and MH method applied to the 1D flow test case. The second column indicates the computational budgets used for the thresholds (in terms of the total number of forward and quantity of interest simulations); the mean and COV (coefficient of variation) are calculated based on 10 estimates of $\mathbb{P}(\mathcal{R}(\para) \geq T | \data)$ for $T^*$ and $T^{**}$. }
    \label{SMC_tab:diff_SMC2}
    \centering
\begin{tabular}{c c|c c c c|c c c c}
	& \begin{tabular}[x]{@{}c@{}}$T^*$/$T^{**}$\\ \footnotesize{$[\times 10^3]$} \end{tabular} & $N$ & $\frac{CESS^{*}}{N}$  & $s$ & $K_R$ & \begin{tabular}[x]{@{}c@{}}Mean $T^*$\\ \footnotesize{$[\times 10^{-3}]$}\end{tabular} & COV $T^*$ & \begin{tabular}[x]{@{}c@{}}Mean $T^{**}$\\ \footnotesize{$[\times 10^{-9}]$}\end{tabular} & COV $T^{**}$   \\
    \hline
    \ssmc & 40/55 & 20 & 0.9 & 20 & 100 & $2.28$ & 0.71 & $3.88$ & 1.72 \\
    \ssmc & 400/550 & 200 & 0.9 & 20 & 100 & $2.45$ & 0.27 & $4.81$ & 0.35 \\
    \ssmc & 510/550 & 20 & 0.9999 & 20 & 100 & $2.60$ & 0.49 & $6.91$ & 1.66 \\
    \ssmc & 400/550 & 20 & 0.9 & 200 & 100 & $2.91$ & 0.65 & $4.77$ & 1.08 \\
    \ssmc & 255/550 & 20 & 0.9 & 20 & 1330 & $2.72$ & 0.44 & $4.31$ & 0.79 \\
    \hline
    MH & 400/550 &-&-&-&-&  $2.45$ & 0.25 & 0 & - 
\end{tabular}
\end{table}

To enable comparison with a basic MH algorithm, we run 10 chains in parallel with one million iterations each. The resulting posterior median and 95$\%$ credible interval of the estimated log-diffusivity are shown in Figure~\ref{SMC_fig:mcmc_diff}a and the resulting samples of $\mathcal{R}(\para) | \data$ in Figure~\ref{SMC_fig:mcmc_diff}b. To visually compare this results with the SMC method, the credible interval in Figure~\ref{SMC_fig:mcmc_diff}a and the particle representation in Figure \ref{SMC_fig:1D_particles}b can be considered. If we would perform MH running three chains in parallel, convergence according to the potential-scale reduction factor ($\hat{R}$-statistics using a target value of 1.2 for all parameters and the second half of the chains; \citeauthor{rstat} \citeyear{rstat}) would be declared after 140'000 iterations and the resulting estimate would be $6.44 \times 10^{-3}$ for $T^*$ and zero for $T^{**}$. This indicates that with the computational budget of the basic version of \ssmc as shown in Table \ref{SMC_tab:diff_SMC2}, we are unable to obtain any reliable estimates with MCMC using MH. With a higher budget of 400,000 for $T^{*}$, the mean of the ten estimates is $2.45 \times 10^{-3}$, and the COV is 0.25. The mean value matches the ones obtained with the \ssmc method. The comparable COV for the same computational budget of \ssmc ($N=200$) is not surprising since the target probability enables enough samples in the MH chains. However, for $T^{**}$, all estimates obtained with MH are zero, even when using the full one million samples per chain.\par

Finally, we would like to highlight the power of including measurement data into this rare event estimation problem. As indicated in Figure~\ref{SMC_fig:1D_adaptive}b, for the prior distribution of the log-conductivity field ($k=0$), $\mathcal{R}(\para) \geq T$ is not a rare event for the considered thresholds. Therefore, we can easily estimate $\mathbb{P}(\mathcal{R}(\para) \geq T)$ under the prior using a limited number of Monte Carlo samples, which gives us 0.23 for $T^{*}$ and 0.22 for $T^{**}$ (here employing 10,000 samples). We conclude that, compared to this previous prior probability of about one quarter, the pumping test measurements lead us to the assessment that the hazard occurrence can be specified as highly unlikely, especially for~$T^{**}$.

\section{2D flow and transport example}
\label{SMC_ssmc_2d}
In the second test case, we infer a hydraulic transmissivity field $\para$ using steady-state pressure data~$\data$ from pumping tests. For the quantity of interest $\mathcal{R}(\para)$, we consider the release of a contaminant on the left side of the model domain and observe the breakthrough of the concentration at a location on the right side of the domain. We are examining a hypothetical scenario where the contamination is expected to no longer pose a risk beyond a pre-defined time frame. That is, the hazard materializes if we observe a breakthrough at the considered location before this time has elapsed.

\subsection{Problem setting}
The aquifer under consideration has a size of $250 \times 250 \times 5$ m and we use a discretization on a grid with $51 \times 51 \times 1$ cells. We assume the properties to be uniform in the vertical direction, thereby simplifying the problem to two spatial dimensions. For the purpose of simulating both the data and the quantity of interest, we utilize the MODFLOW package implemented in Python, specifically the FloPy library \citep{flopy} and 'MT3D-USGS' \citep{bedekar2016mt3d} for the transport simulations.  \par

\begin{figure}	
		\centering
        \includegraphics[width=\textwidth]{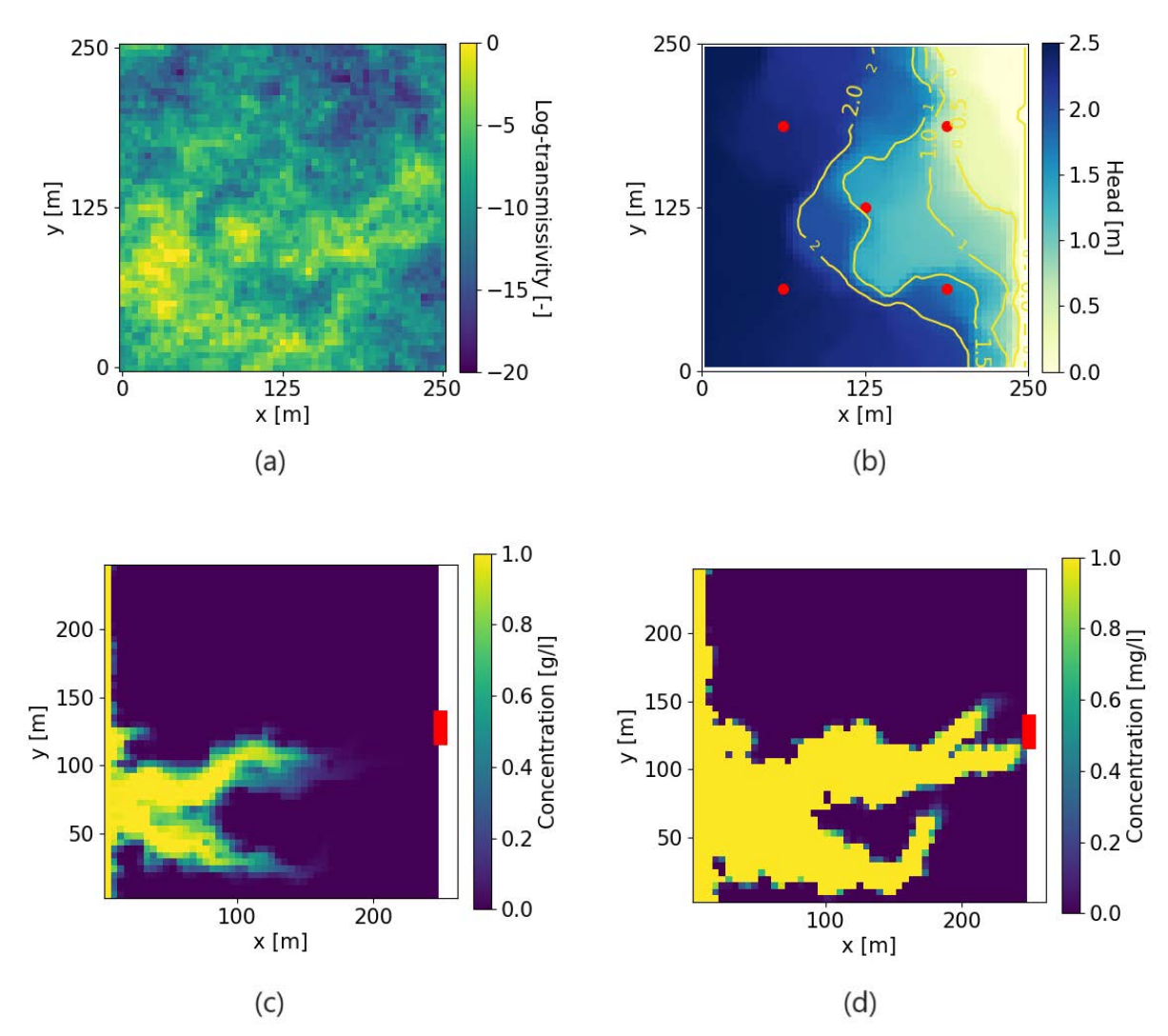}
         % \label{SMC_fig:transm_field}
         % \label{SMC_fig:head_field}
         % \label{SMC_fig:conc_field}
         % \label{SMC_fig:conc_field1}
		\caption{(a) 'True' log-hydraulic transmissivity field and corresponding (b) hydraulic heads resulting from the steady-state pumping test, the red dots indicate the well locations, (c) contamination field and (d) contaminant front after 60 days. }
		\label{SMC_fig:2D_setup}				
\end{figure}

We make the assumption that the system under investigation is confined. The unknown log-transmissivity field $\para$ is assumed to be a Gaussian Random field (\citeauthor{chiles2012geostatistics} \citeyear{chiles2012geostatistics}). We assume a constant mean $\mu_{\log \para} = \log(5 \times 10^{-5})$ with the transmissivity having units of $\mathrm{m}^2/\mathrm{s}$. For the isotropic covariance function, we employ an isotropic exponential covariance function in $\mathbb{R}^2$ with standard deviation $\sigma=3$ and integral scale $l= 25$ m. In order to generate a realization of the $(51\times51)$-dimensional Gaussian random field, we utilize a pixel-based parameterization,
\begin{equation}
\label{SMC_GRF2}
\lat = \boldsymbol{\mu_{\para}} + \boldsymbol{\Sigma_{\para}}^{1/2} \boldsymbol{Z},
\end{equation}
where $\Sigma_{\para}$ denotes the exponential covariance matrix and $\boldsymbol{Z}$ represents a $(51\times51)$-dimensional random vector composed of independent and identically distributed (i.i.d.) standard normal variables. The `true' log-transmissivity field is depicted in Figure~\ref{SMC_fig:2D_setup}a. \par

For the data $\data$, we are considering a five-spot pumping test using a pumping well located in the middle of the model domain and local measurements of the log-transmissivity field at the well locations (Fig.~\ref{SMC_fig:2D_setup}b). For the pumping test, we consider a fixed hydraulic head at the left (2.5 m) and right (0 m) sides of the domain, no-flow boundaries on the other boundaries and pump with a rate of $5 \times 10^{-4}\ \mathrm{m}^3/$s. For the data collection, we consider the steady-state of the system and measure the hydraulic head in four wells centered in the middle of the four quadrants of the domain. For the generation of the synthetic data, we add independent Gaussian observational errors with a standard deviation of 0.02 m. For the local measurements in the five wells, we assume a Gaussian measurement error with a standard deviation of 0.1 (log-scale). Then, we employ standard results for conditional Gaussian random fields, resulting in a mean and covariance matrix in Equation (\ref{SMC_GRF2}), which are conditioned on the local measurements and include their error. \par

For the rare event, we examine a scenario where a contaminant is released on the left side of the model domain, while monitoring the concentration of the contaminant on the right side. Our primary focus lies in determining the time of breakthrough $\mathcal{R}(\para)$ in a critical area in the middle of the right side of the model domain. The hazard is specified as a breakthrough before 60 days ($T = 60$ days), with the breakthrough being specified as the concentration being higher than or equal to 1 mg/l. To simulate this, we assume a constant concentration of 1 g/l on the left side, along with a fixed hydraulic head difference of 2.5 m between the left and right sides (as for the data collection). Additionally, we maintain a constant porosity of 0.3, an effective molecular diffusion coefficient of $10^{-9}$ m$^2$/s, a longitudinal dispersivity of 1 m, a ratio of the transverse to the longitudinal dispersivity of 0.1. Figure~\ref{SMC_fig:2D_setup}c illustrates the concentration distribution after 60 days from the start of the injection for the true field, and Figure~\ref{SMC_fig:2D_setup}d visualizes the corresponding contaminant front.  \par

\subsection{Results}
\begin{figure}	
		\centering
         \includegraphics[width=\textwidth]{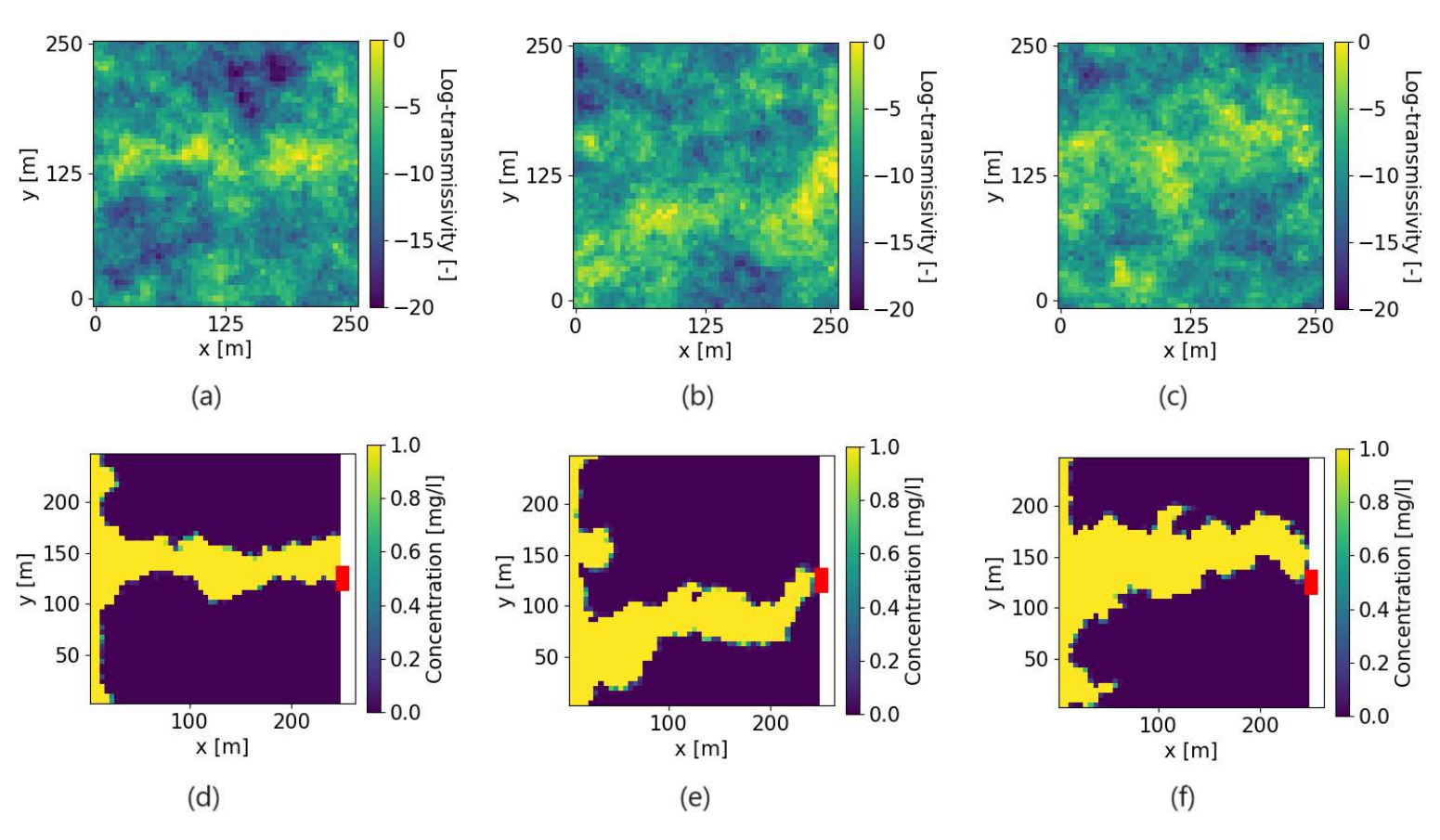}
     \caption{Rare event estimation for the 2D transport example with the \ssmc method (without inversion): (a-c) log-hydraulic transmissivity field examples from the final subset with $\mathcal{R}(\para) \leq 60$~days and (d-f) their corresponding contaminant fronts. }
		\label{SMC_fig:2D_particles}				
\end{figure}

\begin{figure}	
		\centering
         \includegraphics[width=\textwidth]{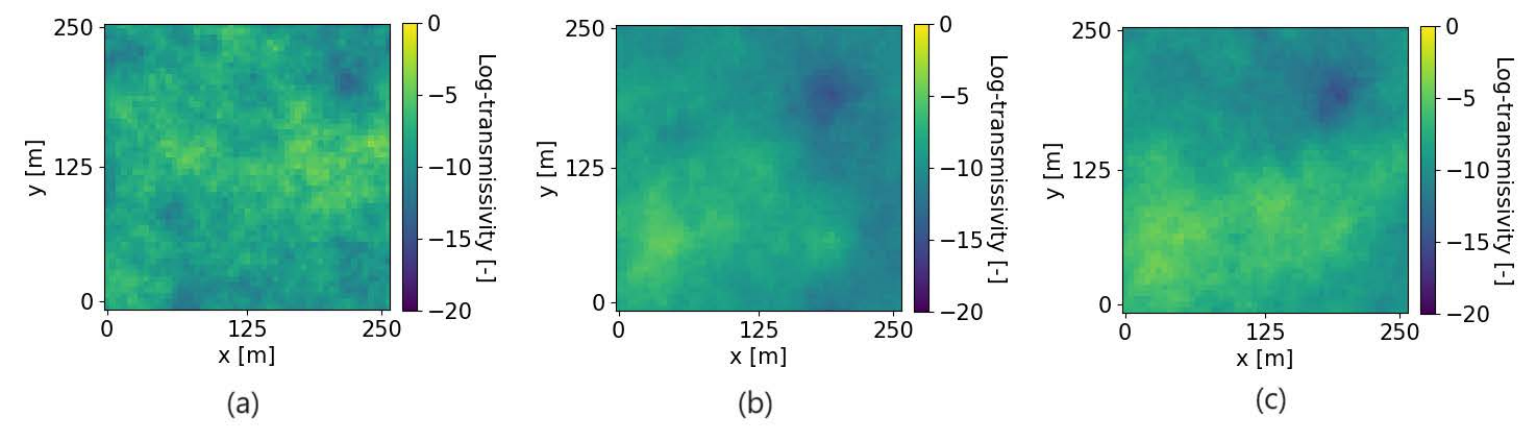}
         % \label{SMC_fig:2D_priormean}
         % \label{SMC_fig:2D_postmean}
         % \label{SMC_fig:2D_postriskmean}
     \caption{Results for the 2D transport example with the \ssmc method: Particle mean representing the log-hydraulic transmissivity field from the (a) prior subset where $\mathcal{R}(\para) \leq 60\ \mathrm{days}$, (b) posterior distribution and (c) posterior subset where $\mathcal{R}(\para) \leq 60\ \mathrm{days}$. }
     \label{SMC_fig:2D_means}				
\end{figure}

We first investigate the occurrence of a contamination breakthrough without incorporating the data. Given the resource-intensive nature of the transport simulations, we adhere to a computational limit of approximately $15,000$ evaluations of $\mathcal{R}(\cdot)$. When using the \ssmc method for this setting, we only employ the second phase and use $N=40$ particles and $s_R=10$ MH steps per subset (Fig. \ref{SMC_fig:flow_rare}). Given that we have demonstrated significant bias when considering an adaptive sequence of thresholds in the one-dimensional flow example (Fig. \ref{SMC_fig:1D_adaptive}), we choose to directly employ a fixed sequence in this test case. We employ a decreasing logarithmic sequence ranging from $T_1=3500$ days down to 100 days, utilizing 30 steps (according to Equation \ref{SMC_log_thres} with $K_P=0$). As the conditional probability during the last steps becomes lower and the risk of the particle system dying is particularly high, we adapt the sequence to steps of five days from 100 days down to the 60 days of interest, leading to $K_R=38$. For the propagation of the particles with MH, we use pCN proposals initialized with a $\rho=1$ (independent proposals), which is then geometrically decreased by a factor of 0.9 in each subset. In this test case, we utilize the pCN proposals as we target a parameter space characterized by high dimensionality (51 $\times$ 51 variables). On the other hand, in the case of the one-dimensional flow example involving only 10 variables, standard Gaussian proposals proved to be effective. In Figure~\ref{SMC_fig:2D_particles}, we provide visual representations of three illustrative log-hydraulic transmissivity field realizations extracted from the final subset where $\mathcal{R}(\para) \leq 60\ \mathrm{days}$. These examples are accompanied by their respective contamination fields. Figure~\ref{SMC_fig:2D_means}a displays the mean transmissivity field of the particles. Running ten repetitions of the \ssmc method, we obtain a mean of $0.71 \times 10^{-4}$ and a COV of 0.37 for $\mathbb{P}(\mathcal{R}(\para) \leq 60\ \mathrm{d})$ (Table \ref{SMC_tab:2D}). With prior sampling and Monte Carlo estimation for the same computational budget, we obtain a mean of $0.87 \times 10^{-4}$ and a COV of 0.60. While the Monte Carlo approach includes zero in the range of the ten probability estimates, the \ssmc method specifies the probability as being at least $0.24 \times 10^{-4}$. \par

\begin{figure}	
		\centering
        \includegraphics[width=\textwidth]{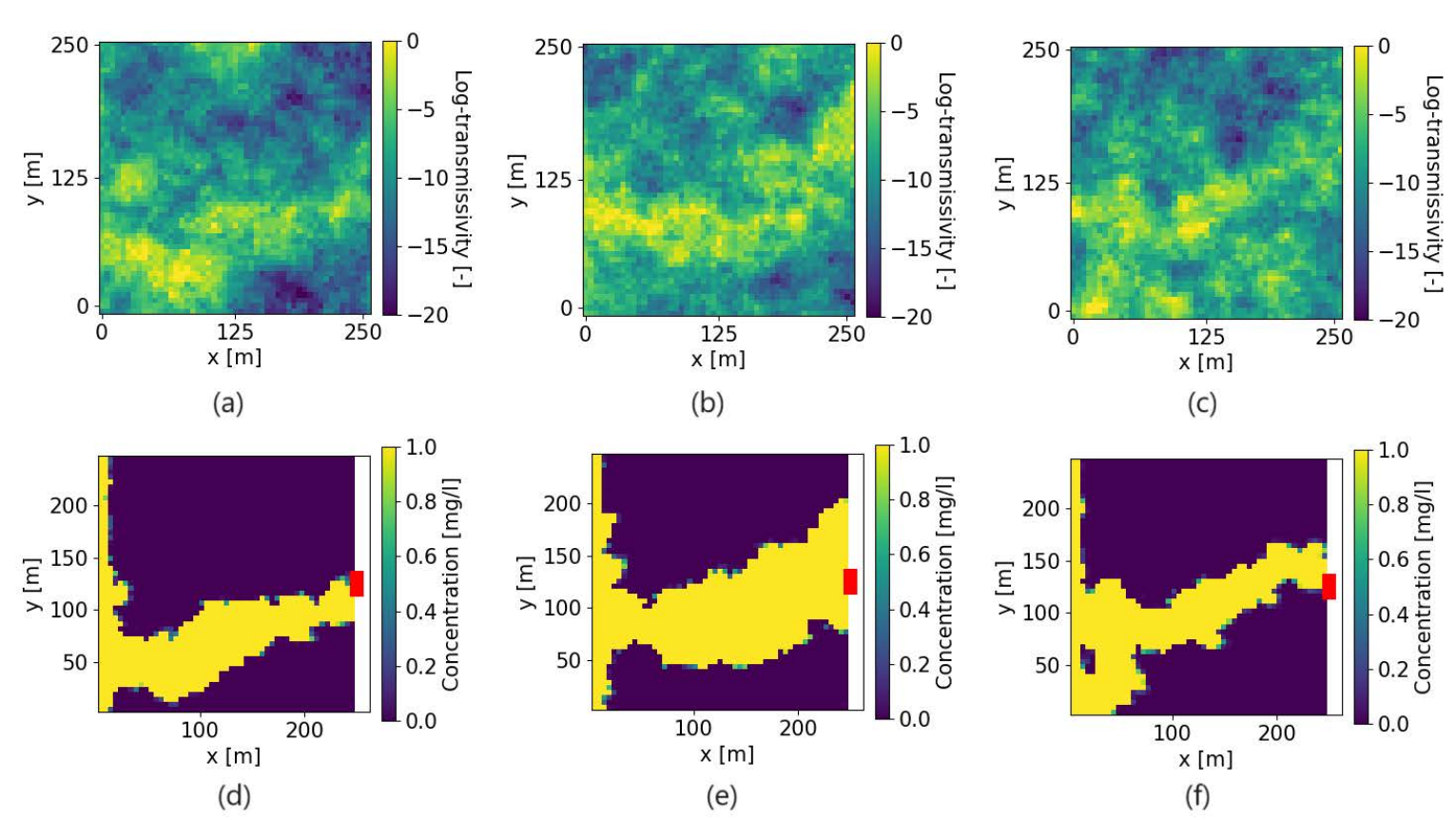}
     \caption{Rare event estimation for the 2D transport example with the \ssmc method (with inversion): (a-c) log-hydraulic transmissivity field examples from the final subset with $\mathcal{R}(\para) \leq 60$~days and (d-f) their corresponding contaminant fronts. }
		\label{SMC_fig:2D_particles_posterior}				
\end{figure}

We now consider the data. Figure~\ref{SMC_fig:2D_setup}d demonstrates that the hazard is occurring for the true log-hydraulic transmissivity field and we are interested to see if the integration of the local and pumping measurements helps to reflect this by increasing the rare event probability estimate. For the posterior inference part of \ssmc, we use a configuration with $N=40$, $CESS^*/N = 0.99$ (leading to $K_P=100$) and $s_P = 100$ MH steps per iteration (Fig. \ref{SMC_fig:flow_post}). A particle estimate of the posterior mean is depicted in Figure \ref{SMC_fig:2D_means}b. For the rare event phase of \ssmc, we implement the adaptation outlined in Section \ref{SMC_dsmc}, wherein we conduct $ss_R=100$ posterior steps within each of the $s_R=10$ MH steps during the rare event phase of the algorithm. This implies that for every subset, we need to assess $\mathcal{R}(\cdot)$ ten times and $\mathcal{G}(\cdot)$ one thousand times. We use the same sequence of thresholds with $K_R=38$ as described above. In total, this results in $N \times (K_P \times s_P + K_R \times s_R \times s_{RR}) = 1.92$ million evaluations of $\mathcal{G}(\cdot)$ and $N \times K_R \times s_R  = 15,200$ evaluations of $\mathcal{R}(\cdot)$ (Table \ref{SMC_tab:2D}). For the propagation, the step size of the pCN proposals is adapted such that the `posterior' steps have an acceptance rate of about 30 $\%$. In Figure~\ref{SMC_fig:2D_particles_posterior}, we showcase three particles from the final posterior subset where $\mathcal{R}(\para) \leq 60\ \mathrm{days}$, along with their corresponding contamination fields. Figure \ref{SMC_fig:2D_means}c shows the mean of the particles lying in the last posterior subset. Upon executing the \ssmc method ten times, we compute an average of $4.56 \times 10^{-4}$ and observe a COV of 0.21 for $\mathbb{P}(\mathcal{R}(\para) \leq 60\ \mathrm{days})$ (Table \ref{SMC_tab:2D}). \par

For a fair comparison with Monte Carlo estimation based on MCMC samples, we run ten chains with 1.92 Million steps and evaluate $\mathcal{R}(\cdot)$ for only 15,000 samples (per chain) that are obtained by thinning. We employ pCN proposals with an adjusted step size aiming for an acceptance rate of 30 $\%$. We obtain a mean rare event probability estimate of $5.64 \times 10^{-4}$ and a COV of 0.49 (Table \ref{SMC_tab:2D}). Using the first three chains, convergence with respect to the $\hat{R}$-statistics would be declared after 350,000 iterations. The corresponding merged 1,500 thinned samples per chain would specify the hazard occurrence probability as zero. \par

\begin{table}[]
\small
    \caption{Table summarizing the different trials of the \ssmc and MH method applied to the 2D transport test case under the prior and the posterior distribution. The second column shows the number of required simulations of the forward response $\mathcal{G}(\cdot)$ and quantity of interest $\mathcal{R}(\cdot)$ and mean, COV (coefficient of variation), min (minimum) and max (maximum) refer to the 10 estimates of the rare event probability. }
    \label{SMC_tab:2D}
    \centering
    \setlength{\tabcolsep}{3pt}
\begin{tabular}{c c c |c c c c |c c c c c c}
	& Method & \begin{tabular}[x]{@{}c@{}}$\mathcal{G}(\cdot) / \mathcal{R}(\cdot)$\\ \footnotesize{$[\times 10^4]$} \end{tabular} & \begin{tabular}[x]{@{}c@{}}Mean\\ \footnotesize{$[\times 10^{-4}]$}\end{tabular} & COV & \begin{tabular}[x]{@{}c@{}}Min \\ \footnotesize{$[\times 10^{-4}]$}\end{tabular} & \begin{tabular}[x]{@{}c@{}}Max \\ \footnotesize{$[\times 10^{-4}]$}\end{tabular} & $N$ & $\frac{CESS^{*}}{N}$  & $s_P$ & $s_R$ & $ss_R$ & $K_R$ \\
    \hline
    Prior & \ssmc & - / 1.5 & 0.71 & 0.37 & 0.24 & 1.09 & 40 & - & - & 10 & - & 38\\
    & Monte Carlo & - / 1.5 & 0.87 & 0.60 & 0 & 1.33 & - & - & - & - & - & - \\
    \hline
    Posterior & \ssmc & 192 / 1.5 & 4.56 & 0.21 & 3.55 & 6.64 & 40 & 0.99 & 100 & 10 & 100 & 38 \\
    & MH & 192 / 1.5 & 5.64 & 0.49 & 2.01 & 12.75 &- &- &- &- &- &-  \\
\end{tabular}
\end{table}

Similar to the one-dimensional flow example, we can observe that incorporating measurements leads to a shift in our estimation of the hazard occurrence probability. In the context of this two-dimensional transport example, the incorporation of local measurements and pumping data increases the estimated probability of hazard occurrence by a factor of about six compared with the estimate based on prior knowledge only. We observe that for the ten considered estimates, the range of the values for the prior and posterior can be clearly separated (for both \ssmc and Monte Carlo estimation). 

\section{Discussion}
\label{SMC_ssmc_disc}
Sustainable groundwater management and assessment of associated hazards are pressing needs that are being accentuated under global change (e.g., \citeauthor{siebert2010groundwater} \citeyear{siebert2010groundwater}, \citeauthor{famiglietti2014global} \citeyear{famiglietti2014global}, \citeauthor{gorelick2015global} \citeyear{gorelick2015global}). With the Posterior Risk Sequential Monte Carlo \ssmcbracket method, we present an approach that combines Bayesian inversion and rare event probability estimation under uncertainty. It first generates a particle approximation of the posterior which is then propagated to provide an accurate estimation of the rare hazard probability. Thereby, the method relies on `subset sampling' and aims to estimate a small probability as a product of larger conditional probabilities. In addition to probability estimation, the method also generates realizations of the rare event (as illustrated in Figs. \ref{SMC_fig:1D_particles} and \ref{SMC_fig:2D_particles_posterior}), providing tangible representations of how the subsurface property field leading to the hazard could look like in practice. In this vein, the \ssmc approach aligns with the perspective of \citet{ferre2017revisiting}, advocating for the communication of information to decision-makers regarding what is known, what is possible, and what remains unknown.\par

In the first phase of the \ssmc method, we employ adaptive SMC for Bayesian inference \citep{zhou2016toward}, relying on power posteriors giving increasingly more weight to the likelihood. The adaptivity of the exponents introduces a slight bias in the results \citep{beskos2016convergence}, but its extent was found to be negligible in the considered test cases. This adaptive feature is attractive as it reduces the number of user-defined tuning parameters and contributes to a more efficient algorithm. The adaptively determined exponents rely on the choice of the target $CESS^*$ (Eq. \ref{SMC_CESS}). The closer this target value is to the number of particles $N$, the better the approximation, but the slower the algorithm becomes as the number of power posteriors grows. The optimal choice of the algorithmic variables $CESS^*$, $s_P$ (number of MH steps) and $N$ (number of particles) depends on the complexity of the posterior distribution, which is influenced by various factors such as the dimension of the parameter space and the underlying physics \citep{amaya2021adaptive}. In their work, \citet{amaya2021adaptive} suggest employing a combination of $CESS^*$ and $s_P$ such that the weighted-mean likelihood of the particles is in agreement with the tempered likelihoods corresponding to the prescribed model throughout the entire run. Moreover, the authors suggest an algorithmic configuration that avoids too frequent resampling steps. To achieve this, they initially set a $CESS^*$ (for instance, $0.99N$), and subsequently fine-tune the number of MH steps ($s_P$) to ensure fitting the data and minimizing the need for resampling. This process can be done employing a smaller number of particles $N$ for preliminary runs, followed by employing a larger number of samples in the 'final' runs. While the accuracy of the approximation improves with an increase in the number of particles $N$, this enhancement comes at a computational cost. However, unlike many MCMC methods, the SMC method is particularly well-suited for parallel computation, as the particles can be distributed across multiple computing nodes. \par
%Generally, as the number of particles increases, the approximation becomes more accurate. Moreover, a slower increase of the exponent and a higher number of MH steps per iteration also contribute to improving the accuracy of the approximation. All three factors come at the expense of computational resources. However, unlike many MCMC methods, the SMC method is particularly well-suited for parallel computation, as the particles can be distributed across multiple computing nodes. The configuration and computational power required ultimately depends on the complexity of the posterior distribution, which is influenced by various factors such as the dimension of the parameter space, the underlying physics, and the characteristics of the measurement setup \citep{amaya2021adaptive}.  \par

In the second phase of the \ssmc method, we rely on subset sampling to estimate the rare event probabilities. The selection of intermediate thresholds involves a trade-off between the intermediate conditional probabilities and the number of particles \citep{au2001estimation}. If the threshold increases slowly, the conditional probabilities are large and a small number of particles is needed to ensure accurate estimation. On the other hand, more intermediate thresholds are needed until the target threshold is reached. If the thresholds increase faster, more particles are needed for an accurate estimation, which also increases the total number of simulations. \citet{cerou2012sequential} propose an adaptive sequence of thresholds based on quantiles to increase the efficiency of their algorithm. The negative aspect of introducing adaptive thresholds is a positive bias in the rare event probability estimate, which diminishes with an increasing number of particles \citep{cerou2012sequential}. \citet{cerou2012sequential} propose a correction factor for the bias, however, their analytical study assumes that the particles are independent, which is hard to guarantee in practice due to the resampling and the finite number of MH steps $s_R$. \par

In the one-dimensional flow example (Section \ref{SMC_ssmc_1d}), the bias resulting from the adaptive thresholds is far from negligible, especially when using a relatively small number of $N=40$ particles and targeting a rare event with probability of one in a billion (Fig.~\ref{SMC_fig:1D_adaptive}). In light of this, we strongly caution against employing an adaptive scheme for the thresholds, particularly if not carefully assessing this bias by re-running with the previously defined sequence as pre-defined thresholds. To avoid bias, and the computational burden associated with re-running or increasing the number of particles, we employ a fixed sequence of thresholds (Eq. \ref{SMC_log_thres}). When the choice of a suitable form for the fixed sequence is unclear, one option is to run an initial adaptive run that can provide valuable insights into appropriate functional forms of the sequence. Similarly, determining the number of subsets $K_R$ can benefit from an initial adaptive run using a ratio of surviving particles guided by the literature (e.g., \citeauthor{cerou2012sequential} \citeyear{cerou2012sequential} recommend 75-80 $\%$). A fixed sequence of thresholds leads to the possibility of the particle system ``dying'' during the rare event estimation process if no particles exceed the current threshold. We did not specifically consider this scenario, but one possible approach to address this issue is discussed by \citet{legland2006sequential}. Their idea involves continuing to generate new particles until a specified count of particles has reached the given threshold. \par
% Both increasing the number of particles $N$ and the number of subsets $K_R$ decreases the risk of a dying particle system. \par

% While increasing the number of particles $N$ seems to be a general recipe to decrease the rare event estimator's variance, increasing $K_R$ leads to fewer particles discarded in each step, which reduces the variance of the estimator. However, it potentially results in high conditional probabilities that are hard to estimate with a small number of particles, which could be a factor mitigating the benefit. Furthermore, in this first simplistic test case with a rather smooth posterior, a slower increase of the exponents (higher $CESS^*$) only had a limited influence on the variance of the rare event estimate. This happens as a small number $s$ of MH steps can prevent the particles from collapsing, even after resampling steps. \par

In the context of the two-dimensional flow and transport example (Section \ref{SMC_ssmc_2d}), posterior exploration presents a challenge as strong non-uniqueness and underdetermination enable a wide range of solutions to accurately explain the observed data \citep{soueid2014hydraulic, cotter2013}. Hence, the number of resampling steps and the propagation through the MH steps play a crucial role in preventing particle collapse. This latter aspect gains even greater significance during the phase of rare event estimation, as resampling cannot be avoided. For this reason, we implement a slight adaptation of the \ssmc method outlined in Figure \ref{SMC_fig:flow1}. Rather than simulating both the forward response and the quantity of interest at each iteration of the rare event phase, we first perform a sequence of $ss_R$ posterior steps during each of the $s_R$ MH steps. We then consider the last state as proposal from the posterior distribution and decide to accept or reject it depending on whether it lies within the current subset. In scenarios involving contamination simulations, where the computational cost of the contamination simulation typically surpasses that of the data simulation flow model, this strategy enhances particle propagation efficiency while simultaneously decreasing computational demands. We suggest to verify that the chosen values for $N$, $s_R$ and $ss_R$ guarantee the generation of significantly new realizations through the MH steps, thereby preventing particle collapse. Visual inspection of particles at various stages of the algorithm can facilitate this assessment. Similar as for the posterior phase, elevating the number of particles $N$ appears to be a suitable strategy for diminishing the variance of the rare event estimator. This proves advantageous, particularly considering the parallelizability of particles. \par

In both test examples, we investigate the significance of using the posterior instead of the prior PDF to determine the probability of hazard occurrence. In the context of the one-dimensional flow example, we showcase how the introduction of pumping test measurements in this scenario alters a rather likely event into a highly unlikely one. Indeed, the initial occurrence probability of roughly a quarter is after considering the data turned into a probability of one in a billion for $T^{**}$. In the case of the two-dimensional transport example, the situation is reversed: the inclusion of local measurements and pumping data helps in quantifying the probability of hazard occurrence as being six times higher than with prior knowledge alone. The integration of posterior inference serves as a clear demonstration of why it is crucial to design appropriate data acquisition strategies within the realm of risk assessment. Designing appropriate experimental designs for such tasks is a research area on its own, as exemplified by \citet{li2010evaluation} and \citet{nowak2012hypothesis} for hydrological settings.\par

We compare the performance of the \ssmc method with a conventional Monte Carlo approach relying on prior or posterior samples obtained by the MH algorithm. In the one-dimensional flow example (Table \ref{SMC_tab:diff_SMC2}), the estimates obtained with \ssmc align with those of the traditional method for the less rare event. For the more rare event with occurrence probability approaching one in a billion, the Monte Carlo approach fails in simulating the hazardous scenario. The \ssmc method, on the other hand, is able to specify the occurrence probability with a coefficient of variation of 0.35. In the two-dimensional transport example (Table \ref{SMC_tab:2D}), the \ssmc method successfully reduces the coefficient of variation by more than 50 $\%$ compared to Monte Carlo estimation based on MH samples (for the inversion setting). This comparison is established within a scenario where Monte Carlo estimation remains feasible. For rarer events, we anticipate complete failure of Monte Carlo estimation, as showcased by the one-dimensional flow example (Table \ref{SMC_tab:diff_SMC2}). \par

It is worth noting that the two phases of the \ssmc method exhibit different dynamics. While in our 1D flow example, the adaptive procedure for the exponents defining the power posteriors leads to an exponential increase, the sequence of thresholds follows a logarithmic progression. In Section \ref{SMC_ssmc_2d}, we take an initial step in addressing this distinct difference in dynamics by using different numbers of MH steps for the two phases of the method. However, there is considerable potential for further exploration and refinement in this regard. So far, we only dealt with rare sets $A = \{ \para \in \mathbb{R}^P: \mathcal{R}(\para) \in \mathcal{T} \}$ with $\mathcal{T} = [T, \infty)$ or $\mathcal{T} = (-\infty, T]$ for some real number $T$. If we would consider $\mathcal{T} = [T^*, T^{**}]$, one could gradually shrink the interval from both sides. Looking ahead, it could be interesting to incorporate surrogate modeling within the \ssmc method to tackle more complex and realistic problems. Surrogates (e.g. \citeauthor{razavi2012review} \citeyear{razavi2012review}) in this context can serve as simplified models or approximations of the underlying system, allowing for faster evaluations and reducing the computational burden. Moreover, considering alternative approaches for the intermediate steps in both phases could be interesting, such as incorporating a method based on smoothed indicator functions and thermodynamic integration proposed by \citet{xiao2019estimation}, in the second phase. Finally, exploring test cases that do not rely on Gaussian assumptions would be intriguing, as previously undertaken for the posterior part in \citet{amaya2021adaptive} and \citet{amaya2022hydrogeological}.

% Moreover, in addition to applications in actual field settings, it would be interesting to conduct a comparative analysis between the \ssmc method and alternative techniques such as the BUS approach \citep{straub2016bayesian} or updated robust reliability measures \citep{jensen2013use}.  

\section{Conclusions}
\label{SMC_ssmc_conc}
The combination of Bayesian inversion and rare event estimation is very helpful for understanding groundwater hazards and their implications for humans and ecosystems. To overcome the challenges of rare event estimation in an inversion setting, we present a two-stage formulation of Sequential Monte Carlo, denoted as the \ssmc method. First, particles are generated to approximate the posterior distribution by adaptively increasing the exponent of the likelihood function. Second, subset sampling is employed to evaluate the probability of the rare event of interest. To showcase the efficacy and accuracy of the \ssmc method, we present a one-dimensional flow example and a two-dimensional flow- and transport example. The one-dimensional example demonstrates that the \ssmc method allows us to estimate rare event probabilities as low as one in a billion. In the two-dimensional example, we showcase the method's capability for rare event probability estimation in a more realistic and complex setting. In both examples, the \ssmc method successfully reduces the coefficient of variation of the rare event probability estimate compared to Monte Carlo estimation based on posterior samples. In both cases, the addition of the measurement data lead to a distinctly different assessment of the occurrence probability than relying on the prior only. Future work will also consider inclusion of surrogate modeling to speed up computations and applications to actual field settings.

\section*{Code availability}
\noindent The code and test examples associated with this article are available in the following GitHub repository: \href{https://github.com/LeaFrie/SMC_groundwater}{https://github.com/LeaFrie/SMC{\_}groundwater}.

\paragraph{Acknowledgments}
We are grateful to David Ginsbourger and Arnaud Doucet for their advice and guidance during this project; their support in discussing methodological questions is greatly appreciated. This work was supported by the Swiss National Science Foundation (project number: \href{http://p3.snf.ch/project-184574}{184574}). 

\section*{References}
\bibliographystyle{plainnat}
\bibliography{references}

\end{document}